\documentclass[12pt]{article}
\usepackage{amsmath}
\usepackage{amsfonts}
\usepackage{float}
\usepackage{amssymb}
\usepackage{latexsym}
\usepackage{graphicx}
\usepackage[english]{babel}
\input epsf.sty

\newcommand{\be}{\begin{equation}}
\newcommand{\ee}{\end{equation}}
\newcommand{\ba}{\begin{array}}
\newcommand{\ea}{\end{array}}
\newcommand{\bqa}{\begin{eqnarray}}
\newcommand{\eqa}{\end{eqnarray}}
\newcommand{\fnl}{f_{\mathrm{NL}}}

\renewcommand{\d}{\mathrm{d}}

\begin{document}

\title{$\alpha$-vacuum and inflationary bispectrum}
 \vspace{3mm}

\author{Wei Xue and Bin Chen\\
{\small Department of Physics} \\
{\small and State Key Laboratory of Nuclear Physics and Technology,}\\
{\small Peking University, Beijing 100871, P.R.China}}

\date{}
\maketitle

\begin{abstract}

In this paper, we discuss the non-Gaussianity originated from the
$\alpha$-vacuum on the CMB anisotropy. For $\alpha$-vacuum,  there
exist correlation 
between points in the acausal two patches of de Sitter spactime.
Such kind of correlation can lead to large local form
non-Gaussianity in $\alpha$-vacuum. For the single field slow-roll
inflationary scenario, the spacetime is in a quasi-de Sitter phase
during the inflation. We will show that the $\alpha$-vacuum in
this case will lead to non-Gaussianity with distinguished feature,
of a large local form and a very different shape.

\end{abstract}

\maketitle
\newpage

\section{Introduction}

 In the standard hot big bang cosmology, there are several tough
 problems,
including the flatness, isotropy and homogeneity, horizon  and
topological defects problems. The hot big bang theory is unable to
answer these problem in a natural way. Inflation, as an add-on, is
remarkably successful in solving these problems. It gives a
natural initial condition for our observed
universe~\cite{Guth81,Linde82,Steinhardt82}. Furthermore the
quantum fluctuations during inflation seed wrinkles in the Cosmic
Microwave Background (CMB), and today's large scale
structure~\cite{Mukhanov81,Guth82,Hawking82,Starobinsky82,Bardeen83}.
As a result, inflation predicts a nearly scale invariant Gaussian
CMB spectrum, which has been confirmed very well in the
experiments\cite{wmap}.

However, inflation as a successful scenario in the very early
universe has its own difficulties. One of the problems with
inflation is that there are too many inflationary models, which
cannot be distinguished by the scalar power spectrum and power
spectrum index from the CMB observation. It is essential to find
more powerful signatures which could distinguish various models
from each other. Moreover, inflation has some conceptual problems,
cosmological singularity problem and Trans-Planckian physics being
two of them. In a sense, inflation scenario is not really a
fundamental theory. The trouble with it mainly comes from our
ignorance of the physics of very early universe, which should be
governed by a quantum gravity theory.

The rapid development of precise experiment cosmology  opens new
windows to very early universe. The scalar spectral index and its
running, the gravitational wave, non-Gaussianity and the
isocurvature perturbation in the CMB~\cite{wmap} are among the
important probes. These probes will not only constrain a large
amount of inflation models and  make the paradigm more clear, but
also shed light on various other issues beyond usual inflation
scenario. These issues include the initial conditions of inflation
model, Trans-Planckian physics and alternative models to
inflation, etc..

Among the probes, non-Gaussianity is one of the most important
ones. It contains much information: magnitude, shape, sign, and
even running. In principle, it could distinguish various inflation
models. The deviation from the Gaussian distribution of the CMB in
the WMAP observation is parameterized by $\fnl$~\cite{fnl}, \be
\zeta= \zeta_g + \frac{3}{5}\fnl \zeta_g^2 \ , \ee where $\zeta$
is the curvature perturbation in the uniform density slices, and
the subscript $g$ denotes the Gaussian distribution. In the WMAP5
year data~\cite{wmap},  two kinds of non-Gaussianity, local form
and equilateral form, have been analysed \be -9 < \fnl^{local} <
111 (95\%\ CL) \ , \quad -151 < \fnl^{equil} < 253 (95 \% \ CL) \
. \ee The central value of the local form non-Gaussianity is $51$.
If the value of the local form non-Gaussianity is confirmed by the
future experiments, such as Planck satellite, then it will be a
great challenge to many inflation models, including the most
well-studied single field inflation models.

In fact, the non-Gaussianity in CMB spectrum may come from various
sources during the evolution. The temperature fluctuation
$\frac{\Delta T}{T}$ is the observable in CMB observation. During
the inflation, the quantum fluctuation of the inflaton $\delta
\phi$ are generated, and the modes of the fluctuation grow with
the exponentially expanding universe. After the fluctuations leave
the horizon, the decoherence effect makes the quantum fluctuations
to be the classical ones. In the large scale, the physical freedom
of scalar perturbation is curvature perturbation $\zeta$. If we
follow a mode, \be {\rm Initial \ Condition \above 0pt (Vacuum) }
\rightarrow \delta\phi \rightarrow \zeta \rightarrow \frac{\Delta
T}{ T} \ .\ee All the transformations are linear at first order,
thus the temperature fluctuations are Gaussian. Meanwhile, any
deviation from linearity in these transformations and the changes
in the initial condition will influence the final observable.
\begin{itemize}

\item  Let us first consider the last stage of the transformation
$\zeta \rightarrow \frac{\Delta T}{ T}$. The fluctuations in the
gravitational potential on the last scattering surface result in
temperature fluctuations in the CMB, which is known as Sachs-Wolfe
effect. The nonlinear Sachs-Wolfe effect generate $\fnl$ of order
one.

\item  The curvature perturbation $\zeta$ is conserved in the
single field inflation, while in the multiple field case, the
entropy perturbation change the evolution of $\zeta$. It will
suppress the perturbation conversion factor for $\zeta$ in this
process. That is why in the curvaton mechanism~\cite{Lyth:2001nq}
and new ekyrotic models~\cite{ekpyrotic} the large local form
non-Gaussianity is possible. (The other important reason for
ekyrotic models generating large non-Gaussianity is that the
slow-roll condition breaks down.)

\item The primordial non-Gaussianity, which resides on the quantum
fluctuation of the scalar field~\cite{NG}, can be from the
microphysics deep in the horizon. Since the observation requires
the potential of the inflaton to be slow-roll,  the interaction of
the inflaton is weak, and non-Gaussianity is only the order of the
slow roll parameter $\fnl \sim {\cal O}(\epsilon, \eta)$. The
picture will change when the modified gravity and non-canonical
action are considered, such as in ghost inflation~\cite{ghost},
DBI inflation~\cite{DBI}, and k-inflation~\cite{K}. The
non-linearity in the action can produce large equilateral form
non-Gaussianity in the CMB. On the other hand, the back reaction
argument \cite{NG} explains why microphysics in the horizon cannot
have large local form non-Gaussianity.

\item The initial condition could be another important source of
non-Gaussianity. One attempt is to consider the thermal vacuum in
the inflationary cosmology~\cite{thermal}, in which the
equilateral and local form non-Gaussianity are both $\gtrsim
\mathcal O (1)$ in some cases. In this paper, we will consider one
parameter family of vacuum states, called $\alpha$-vacuum, in de
Sitter spacetime~\cite{A,dscft} and quasi-de Sitter spacetime in
inflation. In these vacua, there are correlations between points
in the acausal two patches of de Sitter spacetime. We will show
that $\alpha$-vacuum can induce large local form non-Gaussianity.
\end{itemize}

As we know,  the standard treatment in scalar driven inflation
scenario is based on semi-classical gravity, in which the
background is described by classical Einstein Gravity and the
perturbations are quantized in the background. During the slow
roll period, since the evolution of the background is in a
quasi-de-Sitter phase, the perturbations could be treated as the
quantum field in a de-Sitter spacetime. In the expanding
background of inflation, the quantum modes are stretching across
the horizon. The modes observed in the CMB could be in the
trans-Planckian region at the very early time. If the inflation
lasts about 70 e-foldings, the perturbation which we observe in
our Horizon, at that time, is deep inside the horizon. And the
wavelength of these perturbation is smaller than the Planck scale.
The semi-classical description is not applicable for the
perturbation. It is necessary to consider stringy effect or some
other quantum
gravity effects on inflation. 
The trans-Planckian physics in inflation was first raised in
\cite{RHBrev}. As in black hole physics, an efficient way to count
the Trans-Planckian effect is to modify the dispersion relations.
Various dispersion relations and their physical implications have
been
widely studied\cite{MB1}. 
Another way to discuss the trans-Planckian physics is based on the
space-time uncertainty from quantum gravity, such as string
theory. This noncommutative effect will impact the power spectrum
and gravitational waves of CMB~\cite{kempf,HoB} and also modify
the non-Gaussianity\cite{xue2}.


There is another attempt to  address the trans-Planckian physics
in inflation, firstly suggested by Danielsson~\cite{Dan1}. The new
important ingredient in the discussion is the introduction of one
parameter family of $\alpha$-vacuum state  in the inflationary
background. In de Sitter spacetime, the Bunch-Davies vacuum is the
standard vacuum which is invariant under de Sitter space isometry
group. However, due to the absence of globally defined time-like
Killing vector, the vacuum in de-Sitter spacetime cannot be
defined uniquely. Similarly, the choice of the vacuum in the
inflationary background is subtle and may induce observable
signature on CMB data. It was argued that the effective field
theory and semi-classical gravity are applicable from the length
of  new physical scale cutoff to the large scale of the universe.
And it was also assumed that the modes were generated one by one
at the Planck scale or new physics scale $\Lambda$ such that the
initial conditions are imposed at a mode-dependent time instead of
in the infinite past. This is the motivation to introduction of
$\alpha$-vacuum in inflation. Its physical implications on
inflation have been intensely studied.
The order of the correction to the power spectrum has been
discussed
in~\cite{Lowe1,Alberghi,Dan3,Kaloper2,Easther3,npc,Kaloper1,BM4,BM5,Dan4}.
For example, in~\cite{Kaloper1} using the method of  effective
field theory the authors found that the correction was $\sim
\mathcal O (\frac{H^2}{\Lambda^2})$, and in~\cite{npc} the authors
calculated the correction of power spectrum when the modes are
initially created by adiabatic vacuum state, and found that the
correction was $\sim \mathcal O (\frac{H^3}{\Lambda^3})$. The
careful analysis of these different corrections can be seen
in~\cite{BM5}. For the non-Gaussianity from the trans-Planckian
physics it was first roughly analyzed in~\cite{Chen:2006nt}, and
in~\cite{Holman:2007na} its folded form was analyzed in the
effective field theory. In this paper we follow the treatments in
\cite{Dan1,Easther3}. In \cite{Easther3} the authors evaluated the
$\alpha$-vacuum effect in general single field inflation
background, and found an oscillating dependence on the wavenumber
$k$ for the power spectrum. The reason is that during inflation
Hubble scale is not constant, and the coefficient for the
$\alpha$-vacuum state sensitively relies on $k$. We will show that
this $k$-dependence lead to a distinguished feature of
non-Gaussianity.

This paper is organized as follows: in Sec.~\ref{chapter:alpha},
we first discuss the vacuum states in de Sitter space, the
relation between Euclidean vacuum and $\alpha$-vacuum. And then we
introduce different correlation functions, and explain the
property of the antipodal correlation in de Sitter space. In
Sec.~\ref{chapter:ADM} and Sec.~\ref{chapter:three point}, we
review the lagrangian formalism to compute the power spectrum and
three point correlation of the curvature perturbation.
Sec.~\ref{chapter:shape} is the main result of this paper. We
evaluate the local form and equilateral form non-Gaussianity for
Euclidean vacuum and $\alpha$-vacuum  in both de Sitter spacetime
and inflationary backgrounds. We also draw the shapes of
non-Gaussianity in each cases. Finally, we conclude in
Sec.~\ref{chapter:discussion}.

\section{Vacuum state in de Sitter space}
\label{chapter:alpha}

The spacetime of inflation is a quasi-de Sitter spacetime, which
can be conventionally described by the FRW coordinate, \be \d s^2=
\d t^2- e ^{2Ht} \d x^2\ , \label{ds}\ee where $H$ is the Hubble
scale. In this section, to illustrate the feature of
$\alpha$-vacuum clearly, we mainly discuss the de Sitter space in
which $H$ is simply a constant. Note that the metric (\ref{ds})
actually covers only half of de Sitter spacetime.

The equation of motion for a scalar field in the background takes
the form, 
\be \ddot {\delta \phi} + 3H \dot {\delta \phi}-\nabla^2 \delta
\phi+ \frac{\partial V}{\partial \delta \phi}=0\ . \label{eom} \ee
The scalar field could be inflaton, for which the mass of the
scalar field is much less than the Hubble scale $m \ll H$. The
complete solution of (\ref{eom}) can be expressed in momentum
space~\cite{qft}, \be \delta \phi(\tau,
\textbf{x})_k=\frac{1}{2}{\pi^{1/2}H
(-\tau)^{3/2}}e^{i\textbf{k}\cdot \textbf{x}}\Bigg[
c_1(k)\mathrm{H}_\nu ^{(1)}(-k\tau)+c_2(k)\mathrm{H}_\nu
^{(2)}(-k\tau)\Bigg]\ ,\label{phi} \ee where $\tau=-\frac{1}{aH}$
is the conformal time in de Sitter space,\be \nu =
\sqrt{\frac{9}{4} - 12\frac{m^2}{H^2}} \ ,\ee and
$\mathrm{H}_\nu^{({1 \above 0pt 2})}$ are Hankel functions of the
first and second kind. In the limit of $\eta \rightarrow -\infty$
for a fixed $k$, which means that a  mode of the scalar field is
deep in the Hubble horizon, \be \mathrm{H}^{( {1 \above 0pt 2}
)}(-k\eta) \to \left(-\frac{2}{\pi k \eta}\right)^{1/2} e^{( {-
\above 0pt +} ) ik\eta}\ ,
 \ee up to some constant phase factor.

When a mode is deep in the horizon, the spatial scale is much
smaller than the Hubble scale so that the curvature effect is
negligible and the scalar field is well described by quantum field
theory in Minkowski space. As $\eta \rightarrow -\infty$, we could
choose the vacuum as in the flat space and positive frequency
modes from Hankel function of the second kind
$\mathrm{H}_\nu^{(2)}$, i.e. $c_2(k)=0$. This is called thermal
vacuum or Euclidean vacuum in de Sitter spacetime.

The scalar field is quantized by the canonical method, \be \delta
\phi=\sum_n (\delta \phi_n a_n+\delta \phi^*_n a_n^\dag)\ , \ee
where $n$ denotes all the quantum numbers of the modes, and
$\{a_n,a_n^\dag \}$ are the annihilation and creation operators
satisfying the commutative relation \be \label{a} [a_n,
a^{\dag}_{m}]=(2 \pi)^3
 \delta_{mn} \ .\ee And the Euclidean vacuum state is defined to be
\be a_n |\Omega\rangle  =0 \ . \ee

Let's consider a mode $k$ as $\eta \rightarrow - \infty$. The
physical wavelength of the modes is smaller than the Planck scale.
It is natural to set a physical cutoff for momentum $p_c$. When $p
> p_c$, one has to consider the trans-Planckian effect. The
influence of new degrees of freedom and new physical law could be
effectively encoded in the change of dispersion
relation~\cite{MB1}, or space-time
noncommutativity~\cite{kempf,HoB} or some other ways. When
$p<p_c$, the solution of the Klein-Gorden equation is reliable,
the solution of the scalar field is the linear combination of
$\delta \phi(\eta, \textbf{x})_k^{\pm}$ in (\ref{phi}).
A new set of modes for trans-Plankian effect is expressed as the
combination of the Euclidean modes by a Bogoliubov transformation
\cite{M,A} (Mottola-Allen transform), \bqa \tilde \delta \phi_n
&\equiv& N_ \alpha (\delta \phi_n +e^\alpha \delta
\phi^{*}_n)\ ,\nonumber\\
  N_\alpha &&= {1 \over
\sqrt{1-e^{\alpha+\alpha^*}}}\ ,\label{phi*} \eqa where $\alpha$
is a complex number with $\mathrm{Re}\, \alpha<0$ to denote the
rotation of field space, and $N_\alpha$ is derived from the
Wronskian condition or the rule of Bogoliubov transformation.
Since \bqa \delta \phi=\sum_n (\delta \phi_n a_n+\delta \phi^*_n
a_n^\dag)=\sum_n (\tilde \delta \phi_n \tilde a_n+\tilde \delta
\phi^*_n \tilde a_n^\dag)\ ,\nonumber \eqa so the equation of
$\tilde a_n$ can be expressed as \be \tilde a_n = N_\alpha (a_n
-e^{\alpha^*} a_n{}^\dag )\ .\ee Thus the new vacuum called
$\alpha$-vacuum is defined as follows, \be \tilde a_n |\alpha
\rangle=0  \ , \ee where the $\alpha$-vacuum is still de Sitter
invariant, just like Euclidean one. The Bogoliubov transform can
be implemented by a unitary transform \cite{dscft}, \be \tilde
a_n=\mathcal S a_n \mathcal S^{\dag}, \ee where \bqa \mathcal S=
{\rm exp} \left\{ \sum_n c \, (a_n{}^\dagger)^2 - \bar{c} \,
(a_n)^2\right\},~~~~~ c(\alpha) = {1\over 4}\left(\ln\tanh {-{\rm
Re}\, \alpha \over 2} \right) e^{-i\, {\rm Im}\, \alpha}\ .\eqa
The relation between the two vacua is \be |\alpha\rangle =
\mathcal{S} |\Omega\rangle \ .\ee


Let us turn to the Green functions which are useful studying the
power spectrum and non-Gaussianity. There are several kinds of
Green functions, but all of them can be expressed by the Wightman
function.

In the Euclidean vacuum, the Wightman function can be expressed as
\be G_E(x,x')=\langle\Omega|\delta \phi(x)\delta
\phi(x')|\Omega\rangle =\sum _n \delta \phi_n(x)\delta
\phi^*_n(x')\ . \label{Ge}\ee For the distance much smaller than
the Hubble scale, $G_E$ takes the form in Minkowski spacetime, \be
G_E(x,x')\sim \frac{1} {(t-t'-i\epsilon)^2-|\vec x-\vec x'|^2} \
.\label{green}\ee Near the light cone, the Green function is
divergent.

\begin{figure}[h]
\centering
\includegraphics[totalheight=2.5 in]{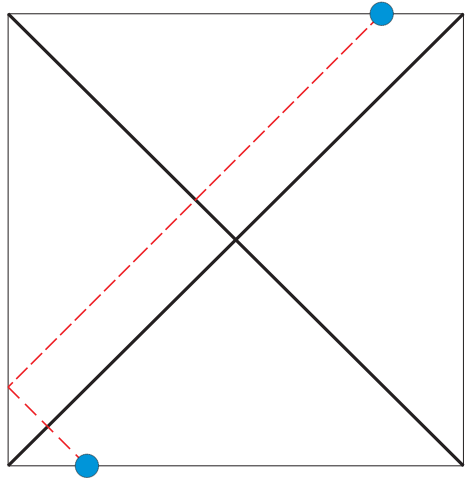}
\caption{The Penrose diagram of de Sitter space, where the two
blue point are antipodal points in the space. The left upper part
of the diagram is the spatial flat patch for a observer on the
left-hand boundary. } \label{fig:de Sitter}
\end{figure}

The metric (\ref{ds}) only covers one half of the whole de Sitter
spacetime. To illustrate the character of Wightman function for
$\alpha$-vacuum, we must extend the analysis to the whole de
Sitter spacetime. The de Sitter spacetime can be constructed as a
hyperboloid in the five-dimensional flat spacetime, \be
-(X^0)^2+(X^1)^2+(X^2)^2+(X^3)^2+(X^4)^2=\frac{1}{H^2}\ , \ee
where $X^i$ is the coordinate in the flat five-dimensional
spacetime. Thus de Sitter spacetime is a maximal symmetric
spacetime with constant curvature, and its symmetry group is
$O(1,4)$. Define the antipodal point of $X$ as $X_A=-X$. In the
Euclidean vacuum, the modes of scalar field can be chosen to obey
the rule \cite{M,A}, \be \delta \phi_n(x_A)=\delta
\phi^*_n(x),\label{apoint}\ee where $x_A$ denotes the antipodal
point to $x$ in de Sitter space as in fig.~\ref{fig:de Sitter}.

In $\alpha$-vacuum, the Wightman function takes the form, \be
G_\alpha(x,x')=\langle\alpha|\delta \phi(x)\delta
\phi(x')|\alpha\rangle = \sum _n\tilde \delta \phi_n(x)\tilde
\delta \phi^*_n(x')\ . \ee With the equation (\ref{Ge}),
(\ref{apoint}), $G_\alpha$ is of the form \be G_{\alpha
}(x,x')=N_\alpha^2\bigl[ G_{E}(x,x')+e^{\alpha+\alpha^*}
G_{E}(x',x) +e^{\alpha^*} G_{E}(x,x'_A)+e^{\alpha}G_{E}(x_A,x')
\bigr] \ .\ee The Wightman function has some special properties.
Firstly, the Green function contains singularity at antipodal
points $\{x_A,x \}$. The singularity can not be observed because
of the separation by the horizon. Secondly, the correlation is not
acausal, because when one calculate the retarded (advanced) Green
function below, the correlation from acausal patches does not
exist. Finally, the Wightman function contains the correlation
between points in the two patches of de Sitter spacetime. It
brings correction to the power spectrum. And most importantly, it
influences the shape of non-Gaussianity, which makes it much
different from the Euclidean vacuum.

In order to calculate the power spectrum and non-Gaussianity from
$\alpha$-vacuum, it is better to use the Green functions in
momentum
space. 
The power spectrum of scalar field can be read from the two-point
correlator in momentum space, \bqa \langle\alpha|\delta
\phi(\textbf{k},\eta)\delta \phi(\textbf{k}',\eta')|\alpha\rangle
&=& (2 \pi)^3\delta({\bf{k}} - {\bf{k}'}) \tilde {\delta
\phi}_{k}(\eta)
        \tilde\delta \phi^*_{k}(\eta')\nonumber\\
&& \hspace{-2.1cm} =(2 \pi)^3\delta({\bf{k}} - {\bf{k}'})N_
\alpha^2 [\delta \phi_{k}(\eta) +e^\alpha \delta
\phi^{*}_k(\eta)][\delta \phi^*_{k}(\eta') +e^{\alpha^*} \delta
\phi_k(\eta')]\ , \eqa where $\delta \phi_{k}$ is the modes of
scalar field in momentum field. In the Euclidean vacuum, for the
leading order approximation \be \delta \phi_{k}(\eta)=
(-H\eta)(1-\frac{i}{k\eta})\frac{e^{-ik \eta}}{\sqrt{2k}}
\ . \label{phik}\ee When the modes cross the horizon, the quantum
fluctuation is transformed to classical one, and the curvature
perturbation is conserved for large scale. Thus we take the time
$\eta$ at the horizon crossing time, which is a good approximation
to calculate power spectrum in single field case. For $k\eta \ll
1$, \bqa \langle\alpha|\delta \phi(\textbf{k},\eta)\delta
\phi(\textbf{k}',\eta')|\alpha\rangle
   &\sim &(2 \pi)^3\delta({\bf{k}} - {\bf{k}'}) \frac{H^2}{2
k^3}N_ \alpha^2(1+e^{\alpha+\alpha^*}-e^{\alpha}-e^{\alpha^*})
\nonumber\\
         &=&(2 \pi)^3  \delta({\bf{k}} -
         {\bf{k}'})\frac{2\pi^2}{k^3} \mathrm P(k)\ .
         \label{P}
    \eqa
Thus the power spectrum of the scalar field can be written as
\be\mathrm P(k)= \frac{H^2}{(2\pi)^2}
\frac{1+e^{\alpha+\alpha^*}-2\rm{Re}\,
e^{\alpha}}{1-e^{\alpha+\alpha^*}}\ee The leading order correction
of power spectrum comes from the $\rm{Re}\, e^{\alpha}$. From the
Wightman function in coordinate space, it is easy to see that the
contribution is from $G_{E}(x,x'_A)$. Meanwhile, if we use a
physical cutoff to set an initial condition of the modes, the
power spectrum should depend on the cutoff scale $\Lambda$, which
could be the string scale, Planck scale, or others.

Next, let us analyze the retarded Green function to prove that the
antipodal point does not break the causality. The retarded Green
function is defined as \be G_R(x,x')=i\theta(t-t')\langle
\rm{VAC}|[\delta \phi(x),\delta \phi(x')]|\rm{VAC}\rangle, \ee
where \bqa \theta(t)= \left\{
\begin{array}{l@{\hspace{5mm}}l}
                1   & t >0 \\ 0  & t<0 \end{array}\right. . \eqa
In the Euclidean vacuum, the retarded Green function in momentum
space takes the form, \bqa G_{RE}(\eta,\tau) = i(2\pi)^3
\delta({\bf{k}} - {\bf{k}'}) \left\{
            \begin{array}{l@{\hspace{5mm}}l}
                0   & \eta < \tau \\
                \delta \phi_k^*(\tau)  {\delta \phi}_k(\eta) -
                {\delta \phi}_k(\tau) \delta \phi_k^*(\eta) & \eta > \tau
            \end{array} \right. .
    \eqa
And in the $\alpha$-vacuum, the retarded Green function takes the
form, \bqa G_{R\alpha}(\eta,\tau) = i(2\pi)^3 \delta({\bf{k}} -
{\bf{k}'}) \left\{
            \begin{array}{l@{\hspace{5mm}}l}
                0   & \eta < \tau \\
                \tilde \delta \phi_k^*(\tau) \tilde {\delta \phi}_k(\eta) -
                \tilde{\delta \phi}_k(\tau) \tilde \delta \phi_k^*(\eta) & \eta > \tau
            \end{array} \right. .
    \eqa
According to the equation (\ref{phi*}), the Green function can be
expressed by the one in the Euclidean vacuum, \be
G_{R\alpha}(\eta,\tau)= N_\alpha^2[
G_{RE}(\eta,\tau)+e^{\alpha+\alpha^*} G_{RE}^*(\eta,\tau)]\ .\ee
From the above equation, the retarded Green function in
$\alpha$-vacuum does not contain the correlation from the two
patches of de Sitter spacetime, so the causality is kept in the
vacuum.

\section{ADM formalism and curvature perturbation in inflationary background}
\label{chapter:ADM}

During inflation, the Hubble radius is changing slowly and the
spacetime is not exactly a de Sitter spacetime. As usual, we have
slow roll parameters, \bqa   \epsilon &\equiv& - \frac{\dot
H}{H^2}= \frac{\dot\phi^2}{2H^2} \simeq
\frac{1}{2}(\frac{V^{\prime}}{V})^2 \ , \nonumber\\
\eta &=& -\frac{\ddot \phi}{H \dot \phi}+\frac{1}{2}\frac{\dot
\phi^2}{H^2} \simeq (\frac{V^{\prime \prime}}{V})=\frac{V^{\prime
\prime}}{3H^2} \ . \label{slowroll1} \eqa The condition $\epsilon
, \eta \ll 1$ indicates that the velocity and the acceleration of
inflaton is quite small. Despite the small deviation from pure de
Sitter spacetime, one can still define $\alpha$-vacuum. However,
the physical implications of $\alpha$ vacuum to  CMB spectrum are
very different. For example, in de Sitter space the power spectrum
has a constant correction with a magnitude of $\mathcal O
(H/\Lambda)$~\cite{Dan1}, where $\Lambda$ is the scale of the new
physics, while in inflationary background the correction is
dependent of wavenumber $k$, so that the power spectrum oscillates
with $k$~\cite{Easther3}.

For simplicity, we just analyze the single field inflationary
models with canonical action. If there exists only one scalar
field in quasi-de Sitter spacetime, then there is just one
perturbation freedom $\delta\phi$ from scalar field. However,
there are also four scalar perturbation freedom from the metric
$\delta g_{\mu \nu}$. $\delta \phi$ and $\delta g_{\mu \nu}$ do
not decouple for scalar
perturbation. 
The gauge invariance removes two of the scalar degrees of freedom
by time and spatial reparametrizaions $x_i\rightarrow
x_i+\partial_i \epsilon(t,x)$ and $t \rightarrow t+
\epsilon(t,x)$~\cite{MFB92}. The constraints in the action remove
two other freedom. Thus there is only one scalar degree of freedom
left. Thus, we should choose a convenient gauge and discuss the
only physical freedom in single field inflationary model.

In general, the spacetime can be decomposed using ADM
formalism~\cite{ADM}, and the metric takes the form \be d s^2 = -
\mathrm N^2 d t^2 + h_{ij} (d x^i + \mathrm N^i dt)(d x^{j}
+\mathrm N^j d t), \label{ADM_metric} \ee where $h_{ij}$ is the
metric of three dimensional spacial slices. The lapse $\rm N$ and
the shift vector $\rm N_i$ contain the freedom of the scalar
perturbation, such as time reparametrization and spatial
reparametrization. With the $3+1$ decomposition, the extrinsic
curvature of the spacial slice is \be K_{ij}=\rm N
\Gamma^{0}_{ij}=\frac{1}{2\rm N}(\dot h_{ij}-\nabla_i \rm
N_j-\nabla_j \rm N_i)\ ,\ee where $\Gamma^{0}_{ij}$ is the
Christoffel symbol in four dimension spacetime. And the intrinsic
curvature of the spacial slices takes the form as  \be
R^{(3)}=R-K_{ij}K^{ij}-K^2 \ ,\ee where \be K= K^{i}_{i} \ .\ee To
simplify the action, we introduce another parameter $E_{ij} \equiv
\rm N K_{ij}$, so the standard Einstein-Hilbert action can be
written as \be S=\frac{1}{2}\int d^4x \sqrt{h}\Bigg[\rm N
R^{(3)}-2\rm N V+ \rm N^{-1}(E_{ij}E^{ij}-E^2)+\rm N^{-1}(\dot
\phi-\rm N^i \partial_i \phi)^2-\rm N h^{ij}\partial_i \phi
\partial_j \phi \Bigg] \ ,\ee where $h=\rm{det} \, h_{ij}$. And in
the action, there is no time derivative of $\rm N$ or $\rm N_i$,
so they are lagrangian multipliers which can be solved directly as
the constraint equations.

It is convenient to choose the comoving gauge, in which the
inflaton perturbation vanishes in the spatial slices, \be \delta
\phi=0\ , \quad h_{ij}=a^2 e^{2\zeta} \delta_{ij} \ .
\label{gauge}\ee The spacial metric $h_{ij}$ is the
nonperturbative form~\cite{NG,NG2}, and the tensor perturbation is
omitted for considering only scalar perturbation. In this gauge,
$\zeta$ is the physical degree of freedom, which is constant
outside the horizon in single field inflation.

In the comoving gauge, the constraint equation is  \bqa &&
\nabla_i[\rm
N^{-1}(E^i_j-\delta^i_j E)]=0 \ ,\nonumber\\
&& R^{(3)}-2V-\rm N^{-2}(E_{ij}E^{ij}-E^2)-\rm N^{-2}\dot \phi^2=0
\ . \label{constraint} \eqa From these constraints and the
equations of the background, the action can be expanded to second
order, third order and even higher order of $\zeta$. In order to
get the action order by order, we need to expand the $\rm N$ and
$\rm N_i$ first. For the shift vector $\rm N_i$, it can always be
decomposed as \be \rm N_i=\partial_i \psi  +\tilde {\rm N}_i \ee
where $\partial_i \tilde {\rm N}_i=0$, and $\psi$ denotes the
scalar perturbation of metric $g_{0i}$. These lagrangian
multipliers can be decomposed in powers of $\zeta$, \bqa
 \rm N
&=& 1+\alpha_1+\alpha_2+\cdots \ ,
\nonumber\\
 \psi  &=& \psi_1+\psi_2+\cdots \ , \nonumber\\
\tilde {\rm N}_i &=& \tilde {\rm N}_i^{(1)}+\tilde {\rm
N}_i^{(2)}+ \cdots \ , \label{decompose:N}\eqa where the subscript
denotes the order, for example $\alpha_n \sim \mathcal O
(\zeta^n)$. From the constraint equation on $\rm
N_i$~(\ref{constraint}), we could get \be \alpha_1=\frac{\dot
\zeta}{H} \quad \quad \partial^2\tilde {\rm N}_i^{(1)}=0 \ .
\label{alpha1}\ee Using appropriate boundary condition, $\tilde
{\rm N}_i$ can be set to $0$. From the view point of inflationary
perturbation, $\tilde {\rm N}_i$ represents the vector
perturbation of the metric, which vanishes in the boundary. From
the constraint equation on $\rm N$~(\ref{constraint}), using the
equation of $R^{(3)}$ \be R^{(3)}=-2a^{-2}e^{-2\zeta}[(\partial
\zeta)^2+2 \partial^2 \zeta] \ ,\label{R3}\ee and Friedmann
equation of the background, the first order of $\psi$ is given by
\be \psi_1= -\frac{ \zeta}{H}+a^2\frac{ \dot
\phi^2}{2H^2}\partial^{-2}\dot \zeta \ . \label{psi1} \ee To get
the action to the quadratic order of $\zeta$, it is enough to
expand $\rm N$ and $\rm N_i$ to the first order of $\zeta$,
because the second order term in $\rm N$ and $\rm N_i$ will
multiply the zero order of constraint equation which is zero. With
the same reason, to get the cubic action for $\zeta$, we do not
need expand the $\rm N$ and $\rm N_i$ to the cubic order. And the
second order expansion of $\rm N$ and $\rm N_i$ in the cubic
action vanish or reduce to total derivatives. So the action to the
quadratic and cubic order of $\zeta$ can be obtained by
substituting $\rm N, \rm N_i$ to the first order  into the  action
and then expanding the action to the second and the third order of
$\zeta$.

After integrating by parts, the quadratic action of $\zeta$ takes
the form as \be S_2=\frac{1}{2}\int d^4x
\frac{\dot\phi^2}{H^2}\left[a^3 \dot \zeta^2- a (\partial \zeta)^2
\right] \ . \label{S2:last}\ee To get the field equation and solve
the curvature perturbation for different modes, a rescale field is
defined as \be v \equiv z \zeta\ , \quad z \equiv a \sqrt{2
\epsilon } \ . \ee The equation of motion is \be
v_k^{\prime\prime} +(k^2-\frac{z^{\prime \prime}}{z}) v_k=0 \ ,
\label{eom:vk}\ee where the prime $\prime$ denotes the derivative
with the conformal time $\tau$. Considering the slow roll
condition the expression of the conformal time is somehow
different from the value in de Sitter space, \be
d\tau=\frac{d(aH)}{(aH)^2(1-\epsilon)} \ . \ee The conformal time
can be written as  \be \tau \simeq -\frac{1}{a
H(1-\epsilon)}\simeq -\frac{1+\epsilon}{a H}  \ . \ee For
power-law inflation, the conformal time takes a exact form, $\tau
= -1/(aH)(1-\epsilon)$. In the field equation, \be \frac{z^{\prime
\prime}}{z} \simeq 2 a^2 H^2
(1+\frac{5}{2}\epsilon-\frac{3}{2}\eta) , \ee so the solution of
(\ref{eom:vk}) is the form of Bessel functions, \be
v_k=\frac{1}{2}(-\pi \tau)^{1/2}\Bigg[ c_1(k)\mathrm{H}_\nu
^{(1)}(-k\tau)+c_2(k)\mathrm{H}_\nu ^{(2)}(-k\tau) \Bigg] \
,\label{solution:vk}\ee where \be \nu=\frac{4}{9}+\epsilon-3\eta \
.\ee The solution is similar to (\ref{phi}). In the case of the
single field inflation, there is only one degree of freedom for
the scalar perturbation, so using scalar field perturbation
$\delta \phi$ as the physical freedom or using the curvature
perturbation, the two kinds of description is the same. On the
large scale, the curvature perturbation $\zeta$ is conserved in
the single field case, while
the scalar field decays to other fields at the end of inflation.
Thus it is more clear to use $\zeta$ as the physical freedom to
describe the perturbation. The results of power spectrum from the
two descriptions are the same up to a factor $2\epsilon$.

When $\tau \rightarrow -\infty $, the Hubble radius is infinite
relative to the modes $k$, the term $\frac{z^{\prime\prime}}{z}$
can be omitted and gravitational effect is negligible, so $v_k
\varpropto e^{-ik\tau}$ in the Euclidean vacuum.
On the other hand, when $k \rightarrow +\infty $, the contribution
to the $v_k$ comes from both the negative and positive frequency
in $\alpha$-vaccum, \be v_k = c_1\frac{
e^{-ik\tau}}{\sqrt{2k}}+c_2 \frac{ e^{+ik\tau}}{\sqrt{2k}}\ .
\label{limit:vk}\ee We may have the question why the equation of
motion is still effective as $k \rightarrow +\infty $. At scale
$\Lambda$, the new physics and some new freedom will emerge. The
new physics scale is set to the Planck scale, string scale or any
other ones. In this paper, we assume that the new physics scale
$\Lambda > H$, and $\Lambda$ is constant. Thus at least on large
scales, the field equation~(\ref{eom:vk}) is reliable. There are
two functions $c_1$ and $c_2$, which depends on the new physics.
What's the most important, we can determine the $c_1(k),c_2(k)$
from the boundary condition at $k/a=\Lambda$. In other words, we
can effectively choose the appropriate boundary condition to take
into account new physics, even without knowing its nature.
Whatever the new physics, it is in the short distance. On the
large scale, the solution just contains the new variable $c_1$ and
$c_2$. The information of the new physics will give different
value of $c_1$ and $c_2$~\cite{Starobinsky:2001kn}. This change of
initial condition will eventually show up in the CMB anisotropy.

Considering  the modes with  wavelength  larger than the new
physics scale, but smaller than the Hubble radius, \be
v_k^{\prime} = -i \sqrt{\frac{k}{2}} c_1
e^{-ik\tau}+i\sqrt{\frac{k}{2}} c_2 e^{+ik\tau} \ .\ee With the
limit value of $v_k$~(\ref{limit:vk}) and $v_k^{\prime}$, and the
boundary condition at $k/a=p_c=\Lambda$, $c_1$ and $c_2$ can be
determined, \bqa
c_1&=&\frac{\sqrt{2k}}{2}e^{ik\tau_c}[v_k(\tau_c)+\frac{i}{k}v_k^{\prime}(\tau_c)]\nonumber\\
c_2&=&\frac{\sqrt{2k}}{2}e^{-ik\tau_c}[v_k(\tau_c)-\frac{i}{k}v_k^{\prime}(\tau_c)]
\ , \label{c1c2:first}\eqa where $\tau_c$ is the time when the k
modes is at the boundary, thus it depends on k, which can be
solved by $k/a=p_c=\Lambda$. As in \cite{Dan1,Easther3} we set the
boundary condition as the wave function at the scale $\Lambda$
only containing emergent wave, and at momentum $p=p_c$, the
quantum fluctuation of scalar field takes the form as \be \frac{d
\delta \phi}{dt}=-ip_c \delta \phi \ . \ee The calculations of
$\delta \phi$ and $\zeta$ in single field inflation are similar,
so at the boundary \be \frac{1}{a} \frac{d
(v_k/a)}{d\tau}=-i\frac{k}{a^2}v_k\ .\label{boundary:D}  \ee With
the relation~(\ref{boundary:D}) and the expression for $c_1$ and
$c_2$~(\ref{c1c2:first}), we could obtain \bqa
c_1&=&\frac{1}{2}[2+i(\frac{Ha}{k})_c]\sqrt{2k}e^{ik\tau_c}v_k(\tau_c)
\nonumber\\
c_2&=&-\frac{i}{2}(\frac{Ha}{k})_c\sqrt{2k}e^{-ik\tau_c}v_k(\tau_c)
 \ ,\label{c1c2:last}\eqa where $(Ha/k)_c=H_c/p_c=H_c/\Lambda$. Note
that in de Sitter space the Hubble scale $H$ is constant, while in
slow roll inflation the Hubble radius is changing $H=H_0
a^{-\epsilon}$. Thus in slow roll inflation $c_1$ and $c_2$ is
dependent of $k$.

In order to compare with the situations of de Sitter space, we
requires $c_2/c_2=e^{\alpha}$ in $\alpha$-vacuum . The solution of
field equation~(\ref{eom:vk}) takes the form
as~(\ref{solution:vk}) up to an unimportant overall phase factor.
Thus the parameter $e^{\alpha}$ is, \be
e^{\alpha}=\frac{c_2}{c_1}=-e^{-2ik\tau_c}\frac{i}{2\frac{\Lambda}{H}+i}
\ ,\ee  from which we know that in de Sitter space $e^{\alpha}$ is
independent of $k$, while in slow roll inflation $e^{\alpha}$ is
dependent of $k$. The magnitude of $e^{\alpha}$ is \be
|e^{\alpha}|=\sqrt{\frac{1}{4\frac{\Lambda^2}{H^2}+1}}\sim
\frac{H}{2\Lambda}\ . \ee

The scalar power spectrum is \be {\rm P}_\zeta =
\frac{k^3}{2\pi^2} |\frac{v_k}{z}|^2 \simeq
\frac{1}{8\pi^2}\frac{H^2}{\epsilon}\frac{1+e^{\alpha+\alpha^*}-2{\rm
Re}\, e^{\alpha}}{1-e^{\alpha+\alpha^*}}(-k\tau)^{3-2\nu}  \ , \ee
where $\nu= \frac{3}{2}+3\epsilon-\eta$. If $\Lambda \gg H$ \bqa
{\rm{Re}}\, e^{\alpha}
&=&-\frac{1}{4\frac{\Lambda^2}{H^2}+1}{\rm{sin}}[2\frac{\Lambda}{H(1-\epsilon)}]+
\frac{2\frac{\Lambda}{H}}{4\frac{\Lambda^2}{H^2}+1}
{\rm{cos}}[2\frac{\Lambda}{H(1-\epsilon)}]  \nonumber\\
&\simeq& \frac{H}{2\Lambda}
{\rm{cos}}[2\frac{\Lambda}{H(1-\epsilon)}] \ ,\label{Reealpha}\eqa
where $H$ is the value when the momentum of modes $k$ is $p_c$. It
is clear that $\frac{\Lambda}{H}$ depends on $k$. The exact
relation can be derived from \be \Lambda=p_c=\frac{k}{a(\tau_c)}\
, \quad H=H_0 a^{-\epsilon}\ , \ee then \be
\frac{\Lambda}{H}\varpropto k^{\epsilon} \ .\ee Therefore, in slow
roll inflation, the correction to the power spectrum will
oscillate with the variable wavenumber $k$. The correction is
$\mathcal O (\frac{H}{\Lambda})$, and the correction is likely to
be observed in the future experiments.

\section{Three-point correlator in Euclidean vacuum}
\label{chapter:three point}

As we discuss in the last section, to get the cubic action of
$\zeta$ we only need to expand the lagrangian multipliers $\rm N$
and $\rm N_i$ to the first order of $\zeta$. The cubic action
could be obtained by substituting the $\rm N$ and $\rm N_i$ to the
ADM action, and expanding the action to the third order of
$\zeta$. Then by integrating by parts many times and using some
technique such as field redefinition, the action can be further
simplified. From the cubic action, the three point correlator is
simply obtained by using path integral formalism at the tree
level.

Integrating by parts and dropping the total derivatives, the cubic
action can be written as~\cite{NG} \bqa
 S_3 &=& \frac{1}{2} \int d^4x a^3
  \Bigg[ \frac{2}{a^2} \frac{\dot{H}}{H^2} \zeta(\partial\zeta)^2 -
  \dot\phi^2  \frac{\dot{\zeta}^3}{H^3}  -
  \frac{4}{a^4}\partial^2\psi_1 \partial_i\zeta \partial_i\psi_{1}
  - \frac{3}{a^4} \zeta \partial^2\psi_1\partial^2\psi_1 \nonumber\\
  \label{vertex:actiona}
  &&\mbox{}  +
  \frac{1}{a^4}\frac{\dot{\zeta}}{H}\partial^2\psi_1\partial^2\psi_1 +
  \frac{3}{a^4}\zeta \partial_i\partial_j\psi_{1} \partial_i\partial_j\psi_{1} -
  \frac{1}{a^4}\frac{\dot{\zeta}}{H}\partial_i\partial_j\psi_{1}\partial_i\partial_j\psi_{1} \Bigg]  .\label{S3:1}\eqa
At first glance, the leading order term in the cubic action is
$\mathcal O(\epsilon^0)$, but after careful integration by parts,
all the terms $\mathcal O(\epsilon^0)$ and $\mathcal
O(\epsilon^1)$ will cancel out, so that the leading order of the
cubic action is $\mathcal O(\epsilon^2)$. If substituting the
equation of $\psi_1$~(\ref{psi1}), after integration by parts, the
action has terms like $\ddot \zeta$. It is convenient to use the
field equation from the quadratic action, \be \frac{\delta
L}{\delta \zeta}|_1=a(\frac{d\partial^2\chi}{dt}+H\partial^2
\chi-\epsilon
\partial^2 \zeta) \ , \label{deltaL}\ee
where \be \partial^2 \chi \equiv a^2 \epsilon \dot \zeta  \ , \ee
$\chi$ is the second term in $\psi_1$~(\ref{psi1}). The final
result of the cubic action is \bqa S_3 = \int d^4x \Bigg[   4 a^5
\epsilon^2H \dot{\zeta}^2\partial^{-2}\dot{\zeta}
+2 f(\zeta)\delta L/\delta \zeta|_1 \Bigg]
 \ ,  \label{S3}\eqa
where \be  f(\zeta)=\frac{-2\eta+3\epsilon}{4}
\zeta^2+\frac{1}{2}\epsilon
\partial^{-2}(\zeta \partial^2 \zeta)+ \cdots    \ . \ee
Here we omit the terms in $f(\zeta)$ which contains the derivative
of $\zeta$ because $\zeta$ is conserved on the large scale, and
any derivative of $\zeta$ has no contribution. In the cubic
action, the contribution from the $f(\zeta)$ terms is obtained
from the redefinition of $\zeta \mapsto \zeta_n + f(\zeta_n)$.
With the redefinition \be S_2[\zeta] \mapsto S_2[\zeta_n] -\int
d^4x 2f(\zeta_n) \delta L/\delta \zeta|_1  \ ,\ee  the second
terms in the cubic action is canceled. When we calculate the
three-point correlator, both the contributions coming from the
cubic $\zeta$ and the contributions from the redefinition should
be taken into account.

The three point correlator could be computed using the path
integral formalism in the interaction picture, \be \langle
\zeta(t,\textbf{k}_1)\zeta(t,\textbf{k}_2)\zeta(t,\textbf{k}_3)
\rangle_{\mathrm{tree}} =i\int^t_{t_0}dt' \langle
[\zeta(t,\textbf{k}_1)\zeta(t,\textbf{k}_2)\zeta(t,\textbf{k}_3),L_3(t')]\rangle
\
  ,\ee 
where $t_0$ is some time that the modes is deep inside the
horizon. The integration can be divided into three parts. The
first part is from the period during which the modes are deep
inside the horizon. In this range, the modes oscillate rapidly, so
the contribution is simply zero. In Euclidean vacuum the value of
$t_0$ is set to $-\infty$, while in $\alpha$-vacuum the value of
$t_0$ is at the boundary for new physics. We assume that $\Lambda
\gg H$, so that $t_0$ in the $\alpha$-vacuum is also deep inside
the horizon. In both situations, the contribution from this part
is zero. The second part is the region well outside the horizon.
Because the value of $\zeta$ is constant, the contribution only
contains the redefinition of $\zeta$. The third region is near the
horizon, where we use the solution of field
equation~(\ref{solution:vk}) and compute the three point
correlator from the path integral.

The leading order contribution to the tree point correlator in
Euclidean vacuum is as follows:

\begin{itemize}

\item \textsl{Contribution from $\dot{\zeta}^2
\partial^{-2}\dot{\zeta}$}.
 We choose $t_0=-\infty$ and $t=0$, which will not influence the
 final results in Euclidean and $\alpha$-vacuum.
  \bqa \langle
\zeta(\textbf{k}_1)\zeta(\textbf{k}_2)\zeta(\textbf{k}_3)
\rangle&=& - i  (2\pi)^3 \delta(\sum_i \textbf{k}_i)
\zeta_{k_1}(0)\zeta_{k_2}(0)\zeta_{k_3}(0) \nonumber\\
  &&\int_{-\infty}^0 d\tau \,g
   \frac{d}{d\tau}\zeta_{k_1}^\ast(\tau)\frac{d}{d\tau}\zeta_{k_2}^\ast(\tau)
\partial^{-2}\frac{d}{d\tau}\zeta_{k_3}(\tau)
   \nonumber\\
   && + \mbox{perms}+\mbox{c.c.}
     \ ,\label{threepoint:example}\eqa
     where $K=k_1+k_2+k_3$, $perms$ denotes exchanging
     $k_1,k_2,k_3$, and $c.c.$ represents the complex conjugate of
     the preceding terms. The prefactor $g=4a^3\epsilon^2 H$. The three point correlator from
     $\dot{\zeta}^2 \partial^{-2}\dot{\zeta}$ is
     \bqa &&(2\pi)^3 \delta(\sum_i
\textbf{k}_i) \frac{H^4}{2^4 \epsilon}
  \frac{1}{\prod_i k_i^3} \left( \frac{k_1^2k_2^2}{K}  \right)+ \mbox{perms}+ \mbox{c.c.}\nonumber\\
&=&(2\pi)^7  \delta(\sum_i \textbf{k}_i) (\rm P_\zeta)^2
\frac{1}{\prod_i k_i^3} \epsilon \left( \frac{1}{K}
\sum_{i>j}k_i^2 k_j^2 \right)
   \ .
  \eqa

\item \textsl{The redefinition $\zeta \mapsto \zeta_n+({-2\eta
+3\epsilon}/{4}) \zeta_n^2$}.

The three point correlator from this contribution is \be (2\pi)^7
\delta(\sum_i \textbf{k}_i) (\rm P_\zeta)^2 \frac{1}{\prod_i
k_i^3} \frac{-2\eta +3\epsilon}{8} \sum_i k_i^3 \ee

\item \textsl{The redefinition $\zeta \mapsto
\zeta_n+({\epsilon}/{2})\partial^{-2}(\zeta_n
\partial^2 \zeta_n )$}.

The three point correlator from this contribution is  \bqa
&&(2\pi)^3 \delta (\sum_i \textbf{k}_i)\frac{H^4}{2^4 \epsilon^2}
\frac{1}{\prod_i k_i^3}\frac{1}{k_1k_2^2k_3^3}+\mbox{perms}
\nonumber\\
&=& (2\pi)^7 \delta(\sum_i \textbf{k}_i) (\rm P_\zeta)^2
\frac{1}{\prod_i k_i^3} \frac{\epsilon}{8} \sum_{i\neq j} k_ik_j^2
  \eqa

\end{itemize}

Finally, taking all the contributions into accout, we have the
three-point correlator  in Euclidean vacuum~\cite{NG}, \be \langle
\zeta(\textbf{k}_1) \zeta(\textbf{k}_2)\zeta(\textbf{k}_3) \rangle
= (2\pi)^7 \delta(\sum_i \textbf{k}_i) (\rm P_\zeta)^2
\frac{1}{\prod_i k_i^3} \mathcal A \ ,\label{fnl:euclidean}\ee
where \be \mathcal A= \epsilon \frac{1}{K} \sum_{i>j}k_i^2 k_j^2
+ \frac{-2\eta +3\epsilon}{8} \sum_i k_i^3 + \frac{\epsilon}{8}
\sum_{i\neq j} k_ik_j^2 \ .\label{A}\ee

\section{Non-Gaussianity in $\alpha$-vacuum}
\label{chapter:shape}

In this section, we analyze the shape of non-Gaussianity and
especially its local form in  $\alpha$-vacuum. We consider both
the de Sitter spacetime and quasi-de-Sitter spacetime and show
that the local form non-Gaussianity in quasi-de Sitter case has
distinctive feature.

The  power spectrum and bispectrum are defined as \bqa \langle
\zeta(\textbf{k}_1) \zeta(\textbf{k}_2) \rangle &\equiv& (2 \pi)^3
\delta(\textbf{k}_1+\textbf{k}_2)\frac{2\pi^2}{k_1^3}\mathrm
P_{\zeta}(k_1)\ , \\
\langle \zeta(\textbf{k}_1) \zeta(\textbf{k}_2)
\zeta(\textbf{k}_1) \rangle &\equiv& (2\pi)^3
\delta(\textbf{k}_1+\textbf{k}_2+\textbf{k}_3)B_\zeta(k_1,k_2,k_3)\
. \eqa Non-Gaussianity measures the deviation of CMB power
spectrum from Gaussian distribution, \be
\zeta=\zeta_g+\frac{3}{5}f_{\rm NL}\left({\zeta}_g^2-\langle
{\zeta}_g^2\rangle\right) \ , \label{fnl:defination}\ee in which
$f_{NL}$ characterize the size of non-Gaussianity:  \be
\label{f}\frac{6}{5}f_{\rm NL}=\frac{\prod_i
k_i^3}{\sum_ik_i^3}\frac{B_\zeta}{4\pi^4\mathrm P_\zeta^2} \ .\ee

In Euclidean vacuum, the three-point correlator is given by
(\ref{fnl:euclidean}), (\ref{A}), so that the parameter of
non-Gaussianity is given by \bqa \fnl = \frac{10}{3}
\frac{1}{\sum_ik_i^3}\mathcal A
 \ . \eqa

The calculation of the three point correlator in the above section
can be extended to the $\alpha$-vacuum. One can simply plug the
value of $\zeta_k$ for the $\alpha$-vacuum into the
equation~(\ref{threepoint:example}) to evaluate the three-point
correlator.  To distinguish $\zeta$'s in different vacua, we use
$\tilde \zeta_k$ to denote its value in the $\alpha$-vacuum, \bqa
\langle \tilde\zeta(\textbf{k}_1)
\tilde\zeta(\textbf{k}_2)\tilde\zeta(\textbf{k}_3) \rangle  =
(2\pi)^7 \delta(\sum_i \textbf{k}_i) (\rm P_{\zeta^{\prime}})^2
\frac{1}{\prod_i k_i^3} \mathcal A' \ .\eqa Note that $\rm
P_{\zeta}$ is different in the two backgrounds, and it will give
small correction to $\mathcal A'$. And $\mathcal A'$ contains two
parts. In de Sitter space,
\bqa \mathcal A'^{(ds)}&=& N_\alpha^6 (1+
4\mathrm{Re}\,(e^{\alpha}))\mathcal
A(k_1,k_2,k_3)+\tilde {\mathcal A} ^{(ds)}\nonumber\\
\tilde {\mathcal A}^{(ds)} &=& N_\alpha ^3
\mathrm{Re}\,(e^{\alpha})[ \mathcal
A(-k_1,k_2,k_3)+ \mathcal A(k_1,-k_2,k_3)\nonumber\\
&&+ \mathcal A(k_1,k_2,-k_3)-3\mathcal A(k_1,k_2,k_3)] \ ,
 \eqa where $A(k_1,k_2,k_3)$ is defined in equation~(\ref{A}) and we neglect the higher order
contribution from slow roll parameter and ${\rm Re} \,
e^{\alpha}$. In de Sitter space, ${\rm Re} \, e^{\alpha}$ is
independent of $k$, and $N_\alpha = {1 /
\sqrt{1-e^{\alpha+\alpha^*}}} \sim 1$. In the discussion below the
value of $N_\alpha$ is set to $1$, and
$4\mathrm{Re}\,(e^{\alpha})A(k_1,k_2,k_3)$ is the next order
contribution to the first part of $\mathcal A'$, so we will also
neglect it. In quasi-de Sitter spacetime, \bqa \mathcal
A'^{(q)}&=& \mathcal
A(k_1,k_2,k_3)+\tilde {\mathcal A} ^{(q)}\nonumber\\
\tilde {\mathcal A}^{(q)} &=&  \Bigg[
\rm{Re}\,(e^{\alpha}_{k_1})\mathcal A(-k_1,k_2,k_3)+
\rm{Re}\,(e^{\alpha}_{k_2})\mathcal A(k_1,-k_2,k_3) +
\rm{Re}\,(e^{\alpha}_{k_3}) \mathcal
A(k_1,k_2,-k_3)\nonumber\\
&&-[\rm{Re}\,(e^{\alpha}_{k_1})+
\rm{Re}\,(e^{\alpha}_{k_2})+\rm{Re}\,(e^{\alpha}_{k_3})] \mathcal
A(k_1,k_2,k_3)\Bigg] \ ,
 \eqa where the index $k_1,k_2,k_3$ of $e^{\alpha}$
 denote its dependence of wavenumber and the upper index $(q)$ denotes quasi-de Sitter for short. The value of
 $\rm{Re}\,(e^{\alpha})$~(\ref{Reealpha}) sensitively depends on the
 variable $k$, since $\Lambda/H \gg 1$.

There are two forms of non-Gaussianity, which are of particular
importance in the data analysis of WMAP. One is the equilateral
form and the other is local form non-Gaussianity. The equilateral
form requires $k_1 \sim k_2 \sim k_3$, and the three momentum
vector compose a equilateral triangle. In this case, in Euclidean
vacuum and $\alpha$-vacuum $\fnl$ are the same in the leading
order approximation, \be \fnl^{equil} \simeq
\frac{10}{9}(\frac{23}{8}\epsilon -\frac{3}{4}\eta) \ .\ee In both
case, the equilateral form non-Gaussianity $\fnl \sim \mathcal O
(\epsilon,\eta )$.

The local form non-Gaussianity requires that one of $k_i \ll$ the
other two $k$. For instance, $-k_1+k_2+k_3 \sim k_3$. The three
momentum vectors compose a isoceles triangle, $k_1=k_2 \gg k_3$.
The modes $k_3$ exits horizon much earlier than the other two
modes. In Euclidean vacuum, if we take the limit $k_1=k_2 \gg
k_3$, then \be \fnl^{local}=\frac{5}{3}(3\epsilon -
\eta)=\frac{5}{6}(1-n_s) \ ,\ee where $n_s$ is the spectral index,
\be n_s-1= \frac{d {\rm In } {\rm P}_\zeta }{d {\rm In} k}\ . \ee
In Euclidean vacuum, ${\rm P}_\zeta \simeq H^2/8\pi^2\epsilon$,
and $n_s-1=2\eta-6\epsilon$. This means that the local form
non-Gaussianity in Euclidean vacuum is the same order of
$\epsilon$ and $\eta$.

The local form non-Gaussianity in Euclidean vacuum can be
estimated by the back-reaction method\cite{NG}. The modes $k_3$
leaves the horizon earlier than the other two modes. The effect
from the modes $k_3$ is rescaling the background spacetime. The
scale factor changes $a(t) \rightarrow a(t)e^{\zeta_3}\sim
a(t)(1+\zeta_3)$, so that the coordinates change accordingly
$\delta x= \zeta_3 x$, where $\zeta_3$ is the amplitude of the
$k_3$ modes. The back reaction of the background will impact the
modes deep in the horizon. The wavenumber decreases $\delta
k=-\zeta_3 k$, and the modes will leave the horizon earlier, \be
\delta k = \delta a \cdot H =a H \delta t H \ . \ee According to
the equation $\delta a= a H \delta t$, we get the relation $\delta
t = -\zeta_3/H$. From the definition of
$\fnl$~(\ref{fnl:defination}), the local form non-Gaussianity is
\bqa \fnl^{local}= \frac{5}{3} \frac{\Delta \zeta}{\zeta_g^2}=
\frac{5}{3} \frac{\Delta t \frac{d
\zeta}{dt}}{\zeta_g^2}=\frac{5}{3} \frac{\frac{-\zeta_3}{H}
\frac{d \zeta}{d \rm{In}k}H}{\zeta_g^2} \simeq \frac{5}{6}(1-n_s)\
,\eqa where the last equation uses the definition of power index
$n_s$. The back-reaction method gives the same result as the one
obtained in the direct way, but this method could be applied to
other models. It indicates that the microphysics from inflaton
cannot have large local form non-Gaussianity in Euclidean vacuum.
Even if the action is in non-canonical form, such as DBI
inflation~\cite{DBI} and K-inflation~\cite{K}, the local form
non-Gaussianity is still $1-n_s$ up to a order one constant.

However, it is a totally different story in $\alpha$-vacuum.
Firstly, we consider the situation of de Sitter spacetime, where
${\rm Re} \,e^{\alpha}$ is independent of $k$. The non-Gaussianity
parameter $\fnl$ contains the contributions from $A$ and $\tilde
A^{(ds)}$. The contributions from $A$ is the same as in the
Euclidean vacuum, which is omitted for conciseness. Thus $\fnl$
from $\tilde A^{(ds)}$ is \bqa \tilde f_{\rm NL} &=&
\rm{Re}(\mathrm e^{\alpha}) \frac{10}{3 }\frac{1}{\sum_i
k_i^3}\Bigg[ -\frac{-2\eta +3\epsilon}{8} \sum_i k_i^3 -
\frac{\epsilon}{8} \sum_{i\neq j}
k_ik_j^2 \nonumber\\
&& \hspace{-2.2cm}+  \epsilon \sum_{i < j} k_i^2
k_j^2(\frac{1}{-k_1+k_2+k_3}+\frac{1}{k_1-k_2+k_3}+\frac{1}{k_1+k_2-k_3}-\frac{3}{k_1+k_2+k_3})\Bigg]\
. \label{fnl:alphads}\eqa If we take the local form limit $k_3 \ll
k_1 \sim k_2$, $-k_1+k_2+k_3 \sim k_3$, then \be  f_{\rm
NL}^{local} \simeq \frac{10}{3}\rm{Re}\,(\mathrm
e^{\alpha})\,\epsilon \frac{k_2}{k_3} \ .\ee Although the
prefactor $\rm{Re}\,(\mathrm e^{\alpha})$ is a small quantity
$\sim {\cal O}(\epsilon \frac{H}{\Lambda})$, $k_2/k_3$ can be a
huge number in the CMB window, say $k_{max}/k_{min} \sim 10^6$ for
WMAP data, so that $f_{NL}^{local}$ could be of order one or even
larger in $\alpha$ vacuum.

Secondly, we consider the local form non-Gaussianity in
inflationary background, where $\mathrm e^{\alpha}$ strongly
depends on $k$. The local form non-Gaussianity takes the form as
\be  f_{\rm NL}^{local} \simeq \frac{5}{3}[\rm{Re}\,(\mathrm
e^{\alpha}_{k_1})+\rm{Re}\,(\mathrm e^{\alpha}_{k_2})]\,\epsilon
\frac{k_2}{k_3} \simeq \frac{10}{3}\rm{Re}\,(\mathrm
e^{\alpha}_{k_2})\,\epsilon \frac{k_2}{k_3} \simeq \frac{5}{3}
\frac{H}{\Lambda} {\rm{cos}}[2\frac{\Lambda}{H(1-\epsilon)}]\,
\epsilon \frac{k_2}{k_3}\ .\ee Similar to de Sitter spacetime,
$\fnl^{local}$ is linear in $\epsilon\frac{k_2}{k_3}$ such that a
large local form non-Gaussianity is possible. However since
$\mathrm {Re (e^\alpha)}$ is $k$-dependent, $\fnl^{local}$ has
distinctive feature.


\begin{figure}[h]
\centering
\includegraphics[totalheight=3in]{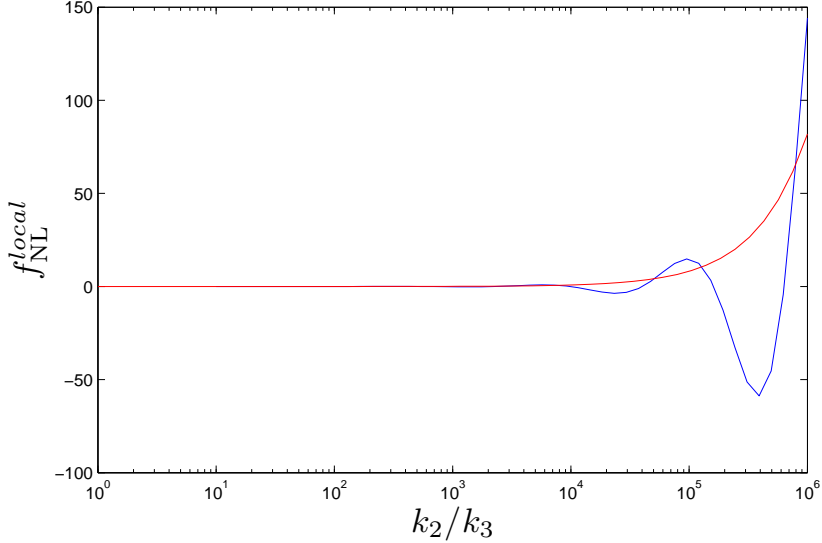}
\caption{The red line represents $\fnl^{local}$ in  de Sitter
space and the blue line represents $\fnl^{local}$ quasi-de Sitter
space } \label{fig:qfnl}
\end{figure}

\begin{figure}[h]
\centering
\includegraphics[totalheight=3in]{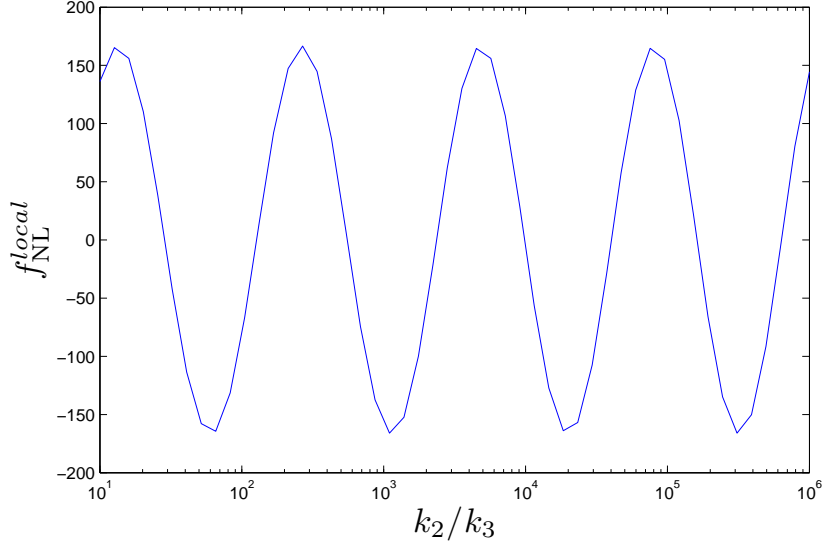}
\caption{$\frac{10}{3}\rm{Re}\,(\mathrm
e^{\alpha}_{k_2})\,\epsilon \times 10^6$, used to compare with
Fig.~\ref{fig:qfnl}} \label{fig:comparef}
\end{figure}

As a example, take $\Lambda$ to be $10^{17} \rm{Gev}$ which is
phenomenological string scale, $H(k_3)=10^{15} \rm{Gev}$, the slow
roll parameter $\epsilon=0.01$, $k_3=1$ (with unit $0.002
Mpc^{-1}$), and $1 \leqslant k_2\lesssim 10^6$. In
Fig.\ref{fig:qfnl}, we draw the local form non-Gaussianity: the
red line represents the one in de Sitter spacetime, while the blue
one represents the one in inflationary background. Obviously, the
local form non-Gaussianity in two cases are different. Especially
in inflationary background, $\rm{Re}\,(\mathrm e^{\alpha}_{k_2})$
is oscillating with modes $k$, as shown in
Fig.~\ref{fig:comparef}.

The possible largeness of local form non-Gaussianity in
$\alpha$-vacuum seems to violate the bound set by the
back-reaction argument. Why the back reaction method is not
applicable in $\alpha$-vacuum? Generally speaking, the correlation
from two patches of de Sitter spacetime makes the back reaction
cannot give the whole effect of non-Gaussianity. In
$\alpha$-vacuum the FRW metric cannot describe the physics
completely, and we must extend the space to the whole de Sitter
space. It is clearer to see the correlation in space coordinates.
For instance, one point $x_1$ is far outside the horizon, and the
other two points $x_2,x_3$ are deep in the horizon. In the whole
de Sitter space, the antipodal point $x_{1A}$ is approaching the
points $x_2,x_3$ deep in the horizon. When $x_{1A}$ is near the
light cone of $x_2$ or $x_3$, there is divergence in tree-level
three-point correlator.

If we carefully analyze $\fnl$ in
$\alpha$-vacuum~(\ref{fnl:alphads}), we will find that there is
also divergence for folded form non-Gaussianity.\footnote{The
folded non-Gaussianity in $\alpha$-vacuum is also discussed in
\cite{Chen:2006nt}, in which they deal with
$e^{\alpha}=constant$.} Therefore, it is interesting to determine
the shape of non-Gaussianity, which has potential to distinguish
different inflationary models if data analysis is accurate
enough.\footnote{We would like to thank Xingang Chen for his
careful explanation and discussion about the shape of
non-Gaussianity.}

The definition of the shape is \be \frac{A}{k_1k_2k_3} \  .\ee We
divide $A$ into several parts. We use subscript $\epsilon$ and
$\eta$ to denote the parts proportional to $\epsilon$ or $\eta$.
In any case, we always have the contribution without including the
modifications from $\alpha$-vacuum: \bqa \mathcal A_{\epsilon}&=&
\epsilon\Bigg( \frac{1}{K} \sum_{i>j}k_i^2 k_j^2 + \frac{3}{8}
\sum_i k_i^3 + \frac{1}{8} \sum_{i\neq
j} k_ik_j^2 \Bigg) \\
\mathcal A_\eta &=& \eta  \Bigg( \frac{-1 }{4} \sum_i k_i^3
\Bigg), \eqa which has been discussed in \cite{NG}. The
modifications in de Sitter spacetime are
 \bqa
 \tilde {\mathcal A}_\epsilon^{(ds)} &=&  \epsilon \rm{Re}(\mathrm
e^{\alpha}) \Bigg[ -\frac{ 3}{8} \sum_i k_i^3 - \frac{1}{8}
\sum_{i\neq j} k_ik_j^2
\nonumber\\
&&
+  \sum_{i < j} k_i^2
k_j^2(\frac{1}{-k_1+k_2+k_3}+\frac{1}{k_1-k_2+k_3}
\nonumber\\
&& +\frac{1}{k_1+k_2-k_3} -\frac{3}{k_1+k_2+k_3})\Bigg]
\\
\tilde {\mathcal A}_\eta^{(ds)} &=&  \eta \rm{Re}(\mathrm
e^{\alpha}) \Bigg( \frac{ 1}{4} \sum_i k_i^3 \Bigg). \eqa
 And the
modifications in quasi-de-Sitter case are
 \bqa
 \tilde {\mathcal A}_\epsilon^{(q)} &=&  \epsilon  \Bigg[\frac{1}{3}[\rm{Re}(\mathrm
e^{\alpha}_{k_1})+\rm{Re}(\mathrm
e^{\alpha}_{k_2})+\rm{Re}(\mathrm e^{\alpha}_{k_3})] (-\frac{
3}{8} \sum_i k_i^3 - \frac{1}{8} \sum_{i\neq j} k_ik_j^2)
\nonumber\\
&& 
+  \sum_{i < j} k_i^2 k_j^2(\frac{\rm{Re}(\mathrm
e^{\alpha}_{k_1})}{-k_1+k_2+k_3} +\frac{\rm{Re}(\mathrm
e^{\alpha}_{k_2})}{k_1-k_2+k_3}+\frac{\rm{Re}(\mathrm
e^{\alpha}_{k_3})}{k_1+k_2-k_3}
\nonumber\\
&& -\frac{\rm{Re}(\mathrm e^{\alpha}_{k_1})+\rm{Re}(\mathrm
e^{\alpha}_{k_2})+\rm{Re}(\mathrm
e^{\alpha}_{k_3})}{k_1+k_2+k_3})\Bigg]\\
\tilde {\mathcal A}_\eta^{(q)} &=&  \eta [\rm{Re}(\mathrm
e^{\alpha}_{k_1})+\rm{Re}(\mathrm
e^{\alpha}_{k_2})+\rm{Re}(\mathrm e^{\alpha}_{k_3})]\Bigg( \frac{
1}{12} \sum_i k_i^3 \Bigg) \ ,
 \eqa
which are very different from de Sitter spacetime case.

\begin{figure}
\centering
\includegraphics[totalheight=3.5in]{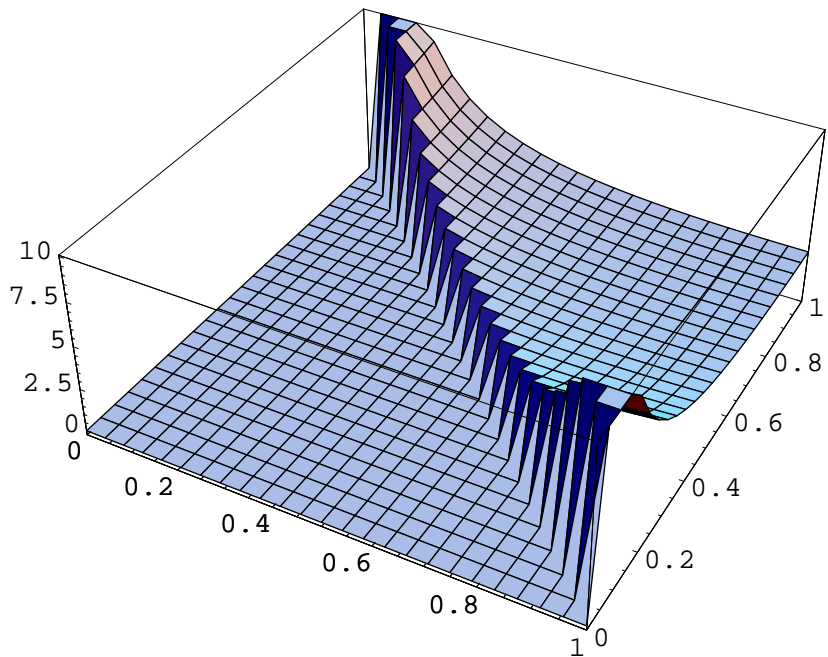}
\caption{$\mathcal A_{\epsilon}/k_1k_2k_3$. This is the shape for
Euclidean vacuum proportional to $\epsilon$. With the shape of
$A_{\eta}/k_1k_2k_3$, they give the leading contribution for
$\alpha$-vacuum.} \label{fig:euclidean}
\end{figure}

\begin{figure}
\centering
\includegraphics[totalheight=3.5in]{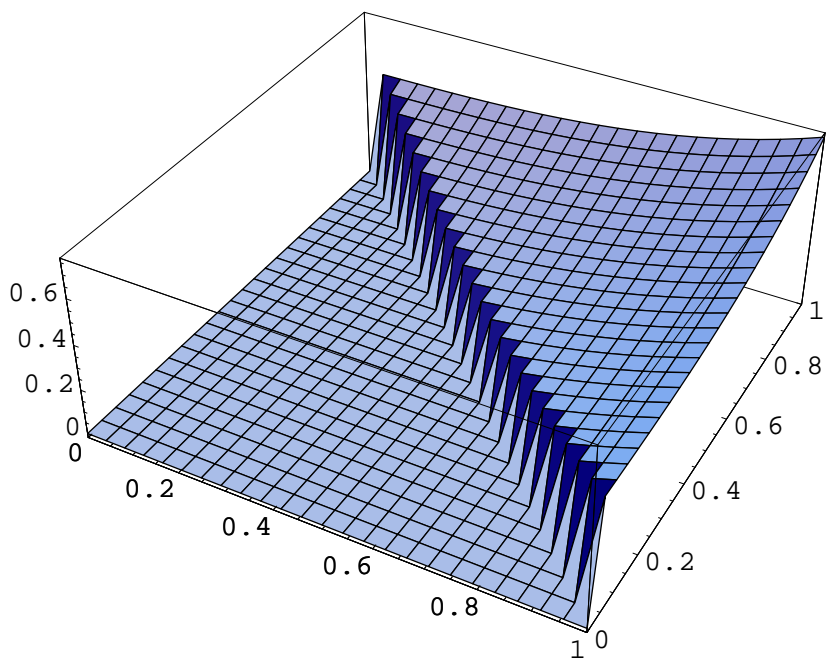}
\caption{$| \mathcal A_{\eta} | /k_1k_2k_3$. This is the shape for
Euclidean vacuum proportional to $\eta$. }\label{fig:eta}
\end{figure}

\begin{figure}
\centering
\includegraphics[totalheight=3.5in]{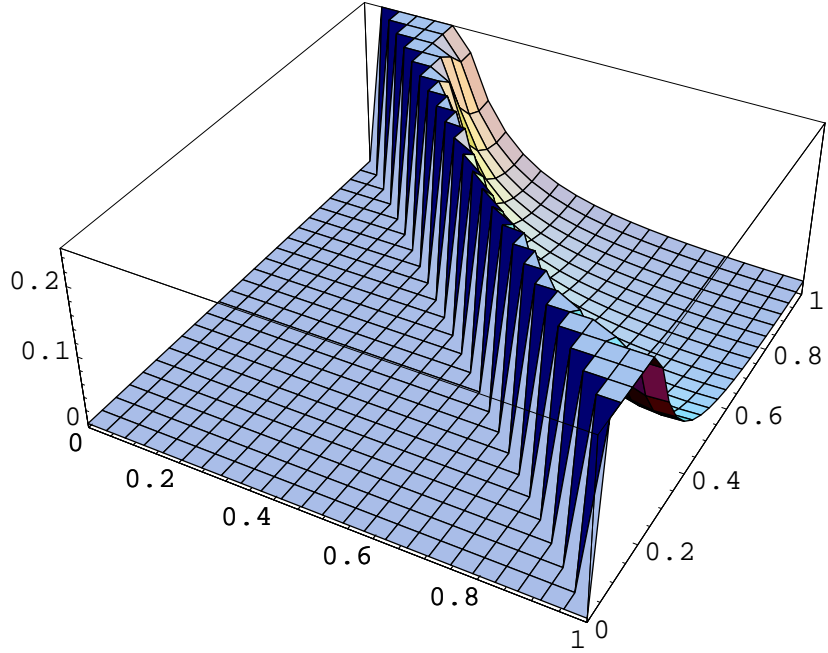}
\caption{$\tilde {\mathcal A}_{\epsilon}^{(ds)}/k_1k_2k_3$. This
is the local $\alpha$-vacuum shape for de Sitter space. The local
form non-Gaussianity near the points (1,0) and (0,1) is large.}
\label{fig:alpha}
\end{figure}

\begin{figure}
\centering
\includegraphics[totalheight=3.5in]{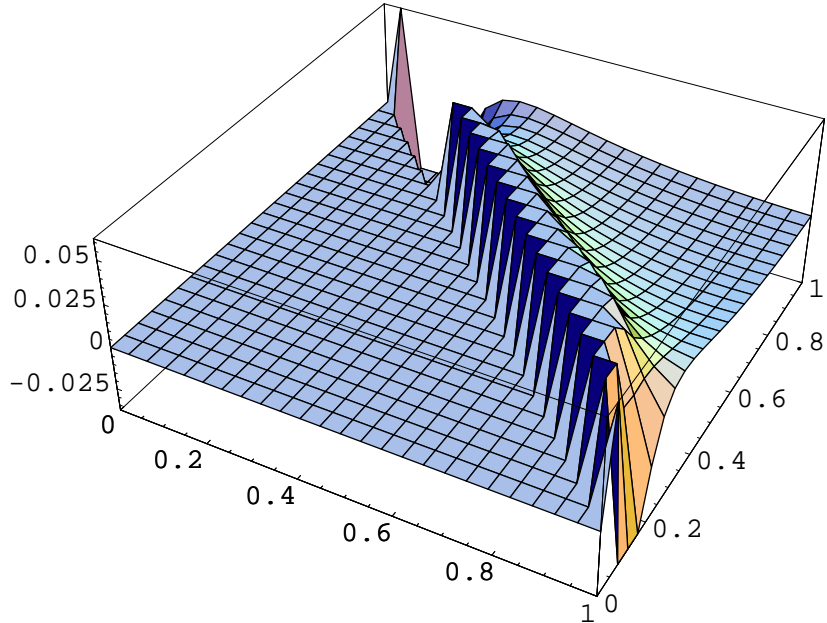}
\caption{$\tilde{\mathcal A}_{\epsilon}^{(q)}/k_1k_2k_3$. This is
the local $\alpha$-vacuum shape proportional to $\epsilon$ for
inflationary space. The local form non-Gaussianity near the points
$(1,0)$ and $(0,1)$ is large, and it reflects some oscillating
character. } \label{fig:qds}
\end{figure}

\begin{figure}
\centering
\includegraphics[totalheight=3.5in]{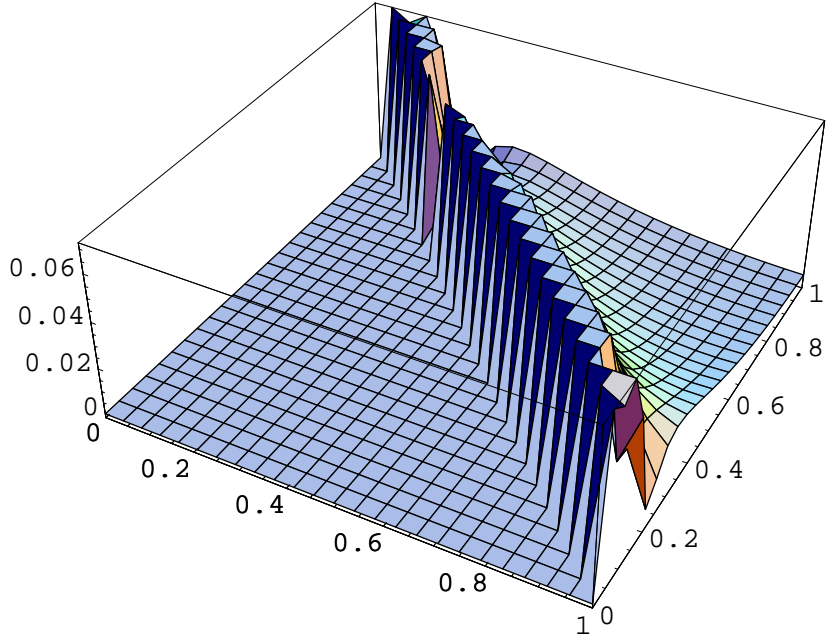}
\caption{$|\tilde{\mathcal A}_{\eta}^{(q)}|/k_1k_2k_3$. This is
the local $\alpha$-vacuum shape proportional to $\eta$  for
inflationary space. The local form is large and oscillating.}
\label{fig:qdsabs}
\end{figure}

 It is more illustrative to draw the shapes of non-Gaussianity in various cases.
  Since $\tilde {\mathcal A}_\eta^{(ds)}$ and $\tilde {\mathcal
A}_\eta^{(q)}$ are small number relative to ${\mathcal A}_\eta$,
we will not draw their shapes. In Fig.~\ref{fig:euclidean}
and~\ref{fig:eta}, we draw the shapes of non-Gaussianity in
Euclidean vacuum. In Fig.~\ref{fig:alpha}, we draw the shape of
$\tilde {\mathcal A}_\epsilon^{(ds)}$, and in Fig.~\ref{fig:qds}
and~\ref{fig:qdsabs}, we draw the shape of $\tilde {\mathcal
A}_\epsilon^{(q)}$. In all the figures, we use the following
convention: $k_3=1$, the x-axis is $k_1/k_3$, the y-axis is
$k_2/k_3$, and the value of z-axis is $\mathcal A / k_1k_2k_3$ up
to a slow roll parameter. The $x-y$ plane diagonal from $(0,1)$ to
$(1,0)$ denotes the folded form, i.e. $k_1+k_2=k_3$. From the
shapes of non-Gaussianity in Euclidean vacuum, the folded form is
finite, except the points $(0,1)$ and $(1,0)$. On contrast, the
shapes of non-Gaussianity in $\alpha$-vacuum, the folded form is
divergent. What's more, the $\fnl$ for the folded form in
$\alpha$-vacuum is divergent as shown in
Equation~(\ref{fnl:alphads}).

The local form non-Gaussianity in the shape is near the point
$(0,1)$ and $(1,0)$. Fig.~\ref{fig:qds} reflects the oscillating
character in inflationary background.

\newpage

\section{Discussion}
\label{chapter:discussion}

In this paper, we studied the physical implication of
$\alpha$-vacuum on the CMB non-Gaussianity. We found that the
$\alpha$-vacuum may lead to  large local form non-Gaussianity, and
its signature 
is distinct. In de-Sitter spacetime, the local form is large which
is proportional to $k_2/k_3$, and in the inflationary background,
the local form could still be large but there is oscillating
character for $\fnl$. Another distinctive feature of
$\alpha$-vacuum is that it leads to a divergent folded form
non-Gaussianity . To illustrate the picture more clearly, we drew
the shapes for different vacua and different backgrounds. We found
that all of these figures have dramatically different feature. For
the de Sitter spacetime the local form is large, and for the
inflationary spacetime the local form not only large, but also
oscillating. These are different from standard slow-roll
inflation, where the local form is proportional to spectral index
of scalar perturbation.

The observational feature from our study is different from
\cite{Chen:2006nt} which only considered the $\alpha$-vacuum
correction from the de Sitter background. 
And even though our study has some overlaps with the one in
\cite{Holman:2007na}, our method and main results are different.
In \cite{Holman:2007na}, they used the effective field theory
(EFT) method during inflation. Since the energy scale of inflation
is very high, from the point of view of EFT, some higher order
derivative terms will be not negligible, and the correction to the
non-Gaussianity origins from the stronger interaction at the
beginning of the inflation. Comparing with trans-Planckian physics
from $\alpha$-vacua, EFT method could be another branch in
inflationary cosmology. Our treatment focused on the influence of
the fluctuation vacuum, while EFT emphasized the action. From the
view of experiments, due to different motivation, the final
prediction of observation is not the same. \cite{Holman:2007na}
analyzed the folded form non-Gaussianities, which has  not been
analyzed in WMAP5 yet. In our study, we concluded that the
trans-Planckian effect would enhance the local form
non-Gaussianities, which is the essential part of the WMAP data
analysis.

There are various interesting issues to address on the implication
of $\alpha$-vacuum on inflation.  First of all, the loop effect in
the $\alpha$-vacuum is still an open question. Steven Weinberg has
given a wonderful discussion about the loop effect in inflationary
correlators~\cite{loop1,loop2,Cogollo:2008bi}. In $\alpha$-vacuum
the problem is focused on how to renormalize the scalar
perturbation in the loop diagram. In some papers, it has been
argued that the $\alpha$-vacuum is not well defined in de Sitter
space due to the divergence~\cite{Banks,Einhorn}. However in
\cite{Lowe2} using the Schwinger-Keldysh formalism a consistent
renormalization method has been constructed to deal with the
divergence.  The other discussion on this issue can be found in
\cite{collins}. It would be interesting to understand the loop
effect and its physical implications in $\alpha$-vacuum.

Secondly, the non-Gaussianity in $\alpha$-vacuum is sensitively
dependent of the initial condition. In the case of single field
inflation with higher derivative terms, the sound speed $c_s$ is
not $1$. For example in the DBI inflation~\cite{DBI} and
K-inflation~\cite{K}, the lagrangian is not canonical, and the
sound speed $c_s \ll 1$ in some situations. The sound horizon $c_s
H^{-1}$ may be smaller than the length scale of new physics. In
this case, the initial condition at the new physics scale is
chosen in the place larger than sound Hubble scale. From the
calculation of the non-Gaussianity, we know that in the path
integral formalism the near Horizon crossing region impacts the
results of non-Gaussianity. The divergence of the non-Gaussianity
in the $\alpha$-vacuum may not exist as pointed out in the
paper~\cite{Chen:2006nt}. This situation also appear in some
special trans-Planckian physics, such as noncommutative
inflation~\cite{HoB}, in the IR region the
effective string scale is smaller than the Hubble scale. 

Thirdly, the different initial condition of $\alpha$-vacuum will
dramatically change the correction of power
spectrum~\cite{npc,Kaloper1}, thus it predicts different
non-Gaussianity. Meanwhile, the trans-Planckian dispersion
relation and noncommutative geometry will also give different
prediction for
non-Gaussianity. 

Fourthly, it is also interesting to consider the physical
implication of the $\alpha$-vacuum in other inflationary models.
One class of them is
 the multiple field inflation~\cite{Nflation}. It has some very different
 signatures from the single field inflation: it has
non-negligible gravitational wave and large local form
non-Gaussianity~\cite{LindeMuk,Bernardeau,Enqvist:2004ey,Gordon:2000hv}.
Another class of inflation models is inspired by string theory. In
particular, DBI inflation is a very remarkable scenario.  It may
give large equilateral non-Gaussianity but no local form
non-Gaussianity. It is worthwhile to discuss the $\alpha$-vacuum
effect in these inflationary models.

Finally, it could be expected that  the tri-spectrum of CMB from
$\alpha$-vacuum is different from the one in Euclidean vacuum. A
more careful investigation would be valuable.

\section*{Acknowledgments}

We would like to thank Robert Brandenberger, Yifu Cai, Xingang
Chen, Eiichiro Komatsu, Yi Wang for valuable discussion. The work
was partially supported by NSFC Grant No. 10535060, 10775002.


\newpage

\begin{thebibliography}{99}


\bibitem{Guth81}
A.~H.~Guth,
 ``The Inflationary Universe: A Possible Solution To The
Horizon And Flatness Problems'', Phys. Rev. {\bf D23}, 347 (1981).

\bibitem{Linde82}
A.~D.~Linde, ``A New Inflationary Universe Scenario: A Possible
Solution Of The Horizon, Flatness, Homogeneity, Isotropy And
Primordial Monopole Problems'',
 Phys. Lett. {\bf B108}, 389 (1982).

\bibitem{Steinhardt82}
A.~Albrecht and P.~J. Steinhardt, ``Cosmology For Grand Unified
Theories With Radiatively Induced Symmetry Breaking'', Phys. Rev.
Lett. {\bf 48}, 1220 (1982).


\bibitem{Mukhanov81} V. ~Mukhanov, and G. ~Chibisov,``Quantum Fluctuation and Nonsingular Universe,''
 JETP {\bf 33}, 549 (1981).

\bibitem{Guth82} A.~ H.~ Guth, and S.~Y.~ Pi, ``Fluctuations in the New Inflationary Universe,''
 Phys. Rev. Lett. {\bf 49}, 1110
(1982).

\bibitem{Hawking82} S. ~W. ~Hawking, ``The Development of Irregularities in a Single Bubble Inflationary Universe,''
 Phys. Lett. {\bf B115}, 295 (1982).

\bibitem{Starobinsky82} A. ~A.~ Starobinsky, `` Dynamics of Phase Transition in the New Inflationary Universe Scenario
 and Generation of Perturbations,'' Phys. Lett. {\bf B117}, 175 (1982).

\bibitem{Bardeen83} J.~ M.~ Bardeen, P. ~J.~ Steinhardt, and M. ~S.
~Turner, ``Spontaneous Creation of Almost Scale - Free Density
Perturbations in an Inflationary Universe,'' Phys. Rev. {\bf D28},
679 (1983).

\bibitem{MFB92}
  V.~ F.~ Mukhanov,  H.~ A.~ Feldman, and  R.~ H.~ Brandenberger,
  ``Theory of cosmological perturbations. Part 1.
  Classical perturbations.
  Part 2. Quantum theory of perturbations. Part 3. Extensions,'' Phys. Rept. {\bf 215}, 203(1992).


\bibitem{wmap}
E.~Komatsu et al., ``Five-Year Wilkinson Microwave Anisotropy
Probe (WMAP) Observations: Cosmological Interpretation,''
submitted to Astrophys. J. Suppl. arXiv:0803.0547 [astro-ph].



\bibitem{fnl}
  E.~Komatsu and D.~N.~Spergel,
  ``Acoustic signatures in the primary microwave background bispectrum,''
  Phys. Rev.  {\bf D 63}, 063002 (2001),
  [arXiv:astro-ph/0005036];
  E.~Komatsu and D.~N.~Spergel,
  ``The cosmic microwave background bispectrum as a test of the physics of
  inflation and probe of the astrophysics of the low-redshift universe,''
  arXiv:astro-ph/0012197;
  E.~Komatsu,
  ``The Pursuit of Non-Gaussian Fluctuations in the Cosmic Microwave
  Background,''
  arXiv:astro-ph/0206039.


\bibitem{qft}
N.~Birrel and P.~Davies, ``Quantum fields in curved space,''
Cambridge Univ. Press 1982.

\bibitem{M}
E.~Mottola, ``Particle Creation In De Sitter Space,'' Phys. Rev.
{\bf D 31}, 754 (1985).

\bibitem{A}
B.~Allen, ``Vacuum States In De Sitter Space,'' Phys. Rev. {\bf D
32}, 3136 (1985).


\bibitem{dscft}
R.~Bousso, A.~Maloney and A.~Strominger, ``Conformal vacua and
entropy in de Sitter space,'' Phys. Rev. {\bf D 65}, 104039
(2002), [hep-th/0112218].


\bibitem{RHBrev}
R.~H.~Brandenberger, Proceeding of the International School on
Cosmology, Kish Island, Iran (2000), [arXiv:hep-ph/9910410].

\bibitem{MB1}
J.~Martin and R.~H.~Brandenberger, ``The trans-Planckian problem
of inflationary cosmology,'' Phys. Rev.  {\bf D 63}, 123501
(2001), [arXiv:hep-th/0005209];
R.~H.~Brandenberger and J.~Martin, ``The robustness of inflation
to changes in super-Planck-scale physics,'' Mod. Phys. Lett. {\bf
A 16}, 999 (2001), [arXiv:astro-ph/0005432];
J.~C.~Niemeyer, ``Inflation with a Planck scale frequency
cutoff,'' Phys. Rev. {\bf D63}, 123502 (2001),
[arXiv:astro-ph/0005533];
J.~C.~Niemeyer, ``Cosmological consequences of short distance
physics,''arXiv:astro-ph/0201511;
J.~Martin and R.~H.~Brandenberger,``A Cosmological window on
transPlanckian physics,'' Proceedings of the Ninth Marcel
Grossmann Meeting on General Relativity, edited by R.~T.~Jantzen,
V.~Gurzadyan and R.~Ruffini, World Scientific, Singapore, 2002,
[arXiv:astro-ph/0012031];
J.~C.~Niemeyer and R.~Parentani, ``Transplanckian dispersion and
scale invariance of inflationary perturbations,'' Phys. Rev. {\bf
D64}, 101301 (2001), [arXiv:astro-ph/0101451];
M.~Lemoine, M.~Lubo, J.~Martin and J.~P.~Uzan, ``The Stress energy
tensor for transPlanckian cosmology,''Phys. Rev. {\bf D65}, 023510
(2002), [arXiv:hep-th/0109128];
R.~H.~Brandenberger, S.~E.~Joras and J.~Martin, ``Trans-Planckian
physics and the spectrum of fluctuations in a bouncing
universe,'' Phys. Rev.  {\bf D 66}, 083514 (2002),
[arXiv:hep-th/0112122].

\bibitem{Starobinsky:2001kn}
  A.~A.~Starobinsky,
  Pisma Zh.\ Eksp.\ Teor.\ Fiz.\  {\bf 73}, 415 (2001)
  [JETP Lett.\  {\bf 73}, 371 (2001)]
  [arXiv:astro-ph/0104043].



\bibitem{Dan1}
U.~H.~Danielsson, ``A note on inflation and transplanckian
physics,'' Phys. Rev.  {\bf D 66}, 023511 (2002),
[arXiv:hep-th/0203198].
U.~H.~Danielsson, ``Inflation, holography and the choice of vacuum
in de Sitter space,'' JHEP {\bf 0207}, 040 (2002),
[arXiv:hep-th/0205227].

\bibitem{Lowe1}
K.~Goldstein and D.~A.~Lowe, ``Initial state effects on the cosmic
microwave background and
 trans-planckian physics,''
Phys. Rev.  {\bf D 67}, 063502 (2003), [arXiv:hep-th/0208167];

\bibitem{Alberghi}
G.~L.~Alberghi, R.~Casadio and A.~Tronconi, ``Trans-Planckian
footprints in inflationary cosmology,'' Phys. Lett. {\bf B 579} ,1
(2004), [arXiv:gr-qc/0303035].

\bibitem{Dan3}
L.~Bergstrom and U.~H.~Danielsson, ``Can MAP and Planck map Planck
physics?,'' JHEP {\bf 0212}, 038 (2002), [arXiv:hep-th/0211006].





\bibitem{Kaloper2}
N.~Kaloper, M.~Kleban, A.~Lawrence, S.~Shenker and L.~Susskind,
``Initial conditions for inflation,'' JHEP {\bf 0211}, 037 (2002),
[arXiv:hep-th/0209231].

\bibitem{Easther3}
R.~Easther, B.~R.~Greene, W.~H.~Kinney and G.~Shiu, ``A generic
estimate of trans-Planckian modifications to the primordial power
spectrum in inflation,'' Phys. Rev.  {\bf D 66}, 023518 (2002),
[arXiv:hep-th/0204129].



\bibitem{npc}
J.~C.~Niemeyer, R.~Parentani and D.~Campo, ``Minimal modifications
of the primordial power spectrum from an
 adiabatic short distance cutoff,''
Phys. Rev.  {\bf D 66}, 083510 (2002), [arXiv:hep-th/0206149].

\bibitem{Kaloper1}
N.~Kaloper, M.~Kleban, A.~E.~Lawrence and S.~Shenker, ``Signatures
of short distance physics in the cosmic microwave  background,''
Phys.  Rev.   {\bf D 66}, 123510 (2002), [arXiv:hep-th/0201158].

\bibitem{BM4}
R.~H.~Brandenberger and J.~Martin, ``On signatures of short
distance physics in the cosmic microwave  background,'' Int. J.
Mod. Phys. {\bf A 17}, 3663 (2002), [arXiv:hep-th/0202142].

\bibitem{BM5}
J.~Martin and R.~H.~Brandenberger `` On the dependence of the
spectra of fluctuations in inflationary cosmology on
transPlanckian physics,''  Phys. Rev. {\bf D68}, 063513(2003),
[arxiv: hep-th/0305161].

\bibitem{Dan4}
U.~H.~Danielsson,`` Transplanckian energy production and slow roll
inflation,''  Phys. Rev. {\bf D 71}, 023516 (2005), [arxiv:
hep-th/0411172].

\bibitem{Holman:2007na}
  R.~Holman and A.~J.~Tolley,
  ``Enhanced Non-Gaussianity from Excited Initial States,''
  JCAP {\bf 0805}, 001 (2008)
  [arXiv:0710.1302 [hep-th]].



\bibitem{Banks}
T.~Banks and L.~Mannelli, ``De Sitter vacua, renormalization and
locality,'' Phys. Rev. {\bf D 67},065009(2003),
[arXiv:hep-th/0209113].

\bibitem{Einhorn}
M.~B.~Einhorn and F.~Larsen, ``Interacting quantum field theory in
de Sitter vacua,'' Phys. Rev.  {\bf D 67}, 024001 (2003),
[arXiv:hep-th/0209159].



\bibitem{Lowe2}
K.~Goldstein and D.~A.~Lowe, ``A note on alpha-vacua and
interacting field theory in de Sitter space,''
[arXiv:hep-th/0302050].

\bibitem{collins} H.~ Collins and R.~
Holman ``Renormalization of initial conditions and the
trans-Planckian problem of inflation,''  Phys. Rev. {\bf D 71}
,085009(2005), [arxiv: hep-th/0501158];
 H.~ Collins and R.~
Holman `` The Renormalization of the energy-momentum tensor for an
effective initial state,''Phys. Rev. {\bf D 74}, 045009(2006),
[arxiv: hep-th/0605107]; D.~ L.~ Nacir and F.~ D. ~Mazzitelli,
``Backreaction in trans-Planckian cosmology: Renormalization,
trace anomaly and selfconsistent solutions,'' Phys. Rev. {\bf D
76} ,024013(2007) ,[ arXiv:0706.2179 [gr-qc]].



\bibitem{kempf}
A.~Kempf, ``Mode generating mechanism in inflation with cutoff,''
Phys. Rev. {\bf D63} (2001) 083514, [arXiv:astro-ph/0009209].
C.~S. ~Chu, B.~R.~ Greene and G.~ Shiu, ``Remarks on inflation and
noncommutative geometry,'' Mod. Phys. Lett. {\bf A 16},2231
(2001),
[hep-th/0011241].
F.~ Lizzi, G. ~Mangano, G.~ Miele and M. ~Peloso, ``Cosmological
perturbations and short distance physics from noncommutative
geometry," JHEP {\bf 0206}, 049 (2002), [arXiv:hep-th/0203099].
A.~Kempf and J.~C.~Niemeyer, ``Perturbation spectrum in inflation
with cutoff,''Phys. Rev. {\bf D64}, 103501 (2001),
[arXiv:astro-ph/0103225].
S.~F.~Hassan and M.~S.~Sloth, ``Trans-Planckian effects in
inflationary cosmology and the modified uncertainty principle,''
 Nucl. Phys. {\bf B 674},434 (2003),
[arXiv:hep-th/0204110];
W.~ Xue, B.~ Chen, and Y.~ Wang, ``Generalized space-time
noncommutative inflation,'' JCAP {\bf 0709}, 011 (2007),
arXiv:0706.1843 [hep-th].


\bibitem{HoB}
R.~ Brandenberger and P.~ Ho, ``Noncommutative space-time, stringy
space-time uncertainty principle, and density fluctuations," Phys.
Rev. {\bf D66} 023517(2002), [arXiv:hep-th/0203119].


\bibitem{xue2}
K.~Fang, B.~Chen and W.~Xue,
  ``Non-commutative Geometry Modified Non-Gaussianities of Cosmological
  Perturbation,''
  Phys.  Rev.  {\bf D 77}, 063523 (2008),
  [arXiv:0707.1970 [astro-ph]];




\bibitem{Lyth:2001nq}
  A.~D.~Linde and V.~F.~Mukhanov,
  ``Nongaussian isocurvature perturbations from inflation,''
  Phys. Rev.   {\bf D 56}, 535 (1997),
  [arXiv:astro-ph/9610219];
  D.~H.~Lyth and D.~Wands,
  ``Generating the curvature perturbation without an inflaton,''
  Phys. Lett.   {\bf B 524}, 5 (2002),
  [arXiv:hep-ph/0110002];
  M.~Li, C.~Lin, T.~Wang and Y.~Wang,
  ``Non-Gaussianity, Isocurvature Perturbation, Gravitational Waves and a No-Go
  Theorem for Isocurvaton,''
  arXiv:0805.1299 [astro-ph];
  Q.~G.~Huang,
  ``Large Non-Gaussianity Implication for Curvaton Scenario,''
  arXiv:0801.0467 [hep-th].





\bibitem{ekpyrotic}
  K.~Koyama, S.~Mizuno, F.~Vernizzi and D.~Wands,
  ``Non-Gaussianities from ekpyrotic collapse with multiple fields,''
  JCAP {\bf 0711}, 024 (2007),
  [arXiv:0708.4321 [hep-th]];
  E.~I.~Buchbinder, J.~Khoury and B.~A.~Ovrut,
  ``Non-Gaussianities in New Ekpyrotic Cosmology,''
  arXiv:0710.5172 [hep-th];
  J.~L.~Lehners and P.~J.~Steinhardt,
  ``Intuitive understanding of non-Gaussianity in ekpyrotic and cyclic
  models,''
  arXiv:0804.1293 [hep-th];
  J.~L.~Lehners and P.~J.~Steinhardt,
  ``Intuitive understanding of non-Gaussianity in ekpyrotic and cyclic
  models,''
  arXiv:0804.1293 [hep-th];
  J.~L.~Lehners and P.~J.~Steinhardt,
  ``Non-Gaussian Density Fluctuations from Entropically Generated Curvature
  Perturbations in Ekpyrotic Models,''
  Phys.\ Rev.\  D {\bf 77}, 063533 (2008),
  [arXiv:0712.3779 [hep-th]];
  J.~L.~Lehners and P.~J.~Steinhardt,
  ``Intuitive understanding of non-Gaussianity in ekpyrotic and cyclic
  models,''
  arXiv:0804.1293 [hep-th].


\bibitem{ghost}
N.~Arkani-Hamed, P.~Creminelli, S.~Mukhoyama, and M.~Zaldarriaga,
``Ghost Inflation,'' JCAP {\bf 0404}, 001 (2004),
[hep-th/0312100].

\bibitem{DBI}
E.~Silverstein and D.~Tong, ``Scalar speed limits and cosmology:
Acceleration from D-cceleration,'' Phys. Rev. {\bf D70}, 103505
(2004), [arXiv:hep-th/0310221]; M.~Alishahiha, E.~Silverstein and
D.~Tong, ``DBI in the Sky,'' Phys. Rev. {\bf D70}, 123505 (2004),
[arXiv:hep-th/0404084];
  X.~Chen,
  ``Multi-throat brane inflation,''
  Phys. Rev.   {\bf D 71}, 063506 (2005),
  [arXiv:hep-th/0408084];
  X.~Chen,
  ``Inflation from warped space,''
  JHEP {\bf 0508}, 045 (2005),
  [arXiv:hep-th/0501184].
X.~Chen,
  ``Running non-Gaussianities in DBI inflation,''
  Phys. Rev.   {\bf D 72}, 123518 (2005),
  [arXiv:astro-ph/0507053];
  M.~Li, T.~Wang and Y.~Wang,
  ``General Single Field Inflation with Large Positive Non-Gaussianity,''
  JCAP {\bf 0803}, 028 (2008),
  [arXiv:0801.0040 [astro-ph]].


\bibitem{K}
J.~ Garriga and V. ~F. ~Mukhanov, ``Perturbations in k-inflation,"
Phys. Lett. {\bf B 458}, 219 (1999), [arXiv:hep-th/9904176];  C.~
Armendariz-Picon, T. ~Damour and V. ~Mukhanov,``k-inflation,"
Phys. Lett. {\bf B 458}, 209 (1999), [arXiv:hep-th/9904075].


\bibitem{Nflation}
S.~ Dimopoulos, S. ~Kachru, J.~ McGreevy and J. ~Wacker,
``N-flation,''[arXiv:hep-th/0507205]; R. ~Easther and L.
~McAllister, ``Random matrices and the spectrum of N-flation,''
 JCAP {\bf 0605} , 018(2006),[arXiv:hep-th/0512102].


\bibitem{LindeMuk}
A.~ Linde and V.~ Mukhanov,`` Nongaussian isocurvature
perturbations from inflation,'' Phys. Rev. {\bf D56}535 (1997),
[arXiv:astro-ph/9610219].

\bibitem{Bernardeau}
F. ~Bernardeau and J. ~Uzan, ``NonGaussianity in multifield
inflation,''Phys. Rev. {\bf D66} (2002) 103506,
[arXiv:hep-ph/0207295].

\bibitem{Enqvist:2004ey}
  K.~Enqvist, A.~Jokinen, A.~Mazumdar, T.~Multamaki and A.~Vaihkonen,
  ``Non-Gaussianity from Preheating,''
  Phys. Rev.  Lett.   {\bf 94}, 161301 (2005),
  [arXiv:astro-ph/0411394].

  \bibitem{Gordon:2000hv}
  C.~Gordon, D.~Wands, B.~A.~Bassett and R.~Maartens,
  ``Adiabatic and entropy perturbations from inflation,''
  Phys. Rev.  D {\bf 63}, 023506 (2001),
  [arXiv:astro-ph/0009131];
\bibitem{Seery:2005gb}
  D.~Seery and J.~E.~Lidsey,
  ``Primordial non-gaussianities from multiple-field inflation,''
  JCAP {\bf 0509}, 011 (2005)
  [arXiv:astro-ph/0506056];
  S.~W.~Li and W.~Xue,
  ``Revisiting non-Gaussianity of multiple-field inflation from the field
  equation,''
  arXiv:0804.0574 [astro-ph];
  X.~Gao,
  ``Primordial Non-Gaussianities of General Multiple Field Inflation,''
  arXiv:0804.1055 [astro-ph].


\bibitem{thermal}
B.~Chen, Y.~Wang, and W.~Xue, ``Inflationary nonGaussianity from
thermal fluctuations'', JCAP {\bf 0805}, 014
(2008),[arXiv:0712.2345 [hep-th]];
 B.~Chen, Y.~Wang, W.~Xue and R.~H.~Brandenberger,
``String Gas Cosmology and Non-Gaussianities'', arXiv:0712.2477
[hep-th].


\bibitem{ADM}
R.~Arnowitt, S.~Deser, and C.~W.~Misner, ``The dynamics of general
relativity,'' gr-qc/0405109.


\bibitem{NG} J.Maldacena, ``Non-Gaussian features of primordial fluctuations
in single field inflationary models," JHEP {\bf 0305}, 013 (2003),
[arXiv:astro-ph/0210603].

\bibitem{NG2} V. Acquaviva, N. Bartolo, S. Matarrese and A. Riotto,
¡°Second-order cosmological perturbations from inflation,¡± Nucl.
Phys. {\bf B 667}, 119 (2003), [arXiv:astro-ph/0209156].
 David Seery and James E. Lidsey, ``Primordial non-gaussianities in single field
inflation" JCAP {\bf 0506}, 003 (2005), [arXiv:astro-ph/0503692].

\bibitem{Chen:2006nt}
  X.~Chen, M.~X.~Huang, S.~Kachru and G.~Shiu,
 ``Observational signatures and non-Gaussianities of general single field
  inflation,''
  JCAP {\bf 0701}, 002 (2007)
  [arXiv:hep-th/0605045].








\bibitem{loop1}
S.~Weinberg, ``Quantum contributions to cosmological
correlations,'' Phys. Rev.  {\bf D72}(2005) 043514,
[hep-th/0506236].

\bibitem{loop2}
S.~Weinberg, ``Quantum contributions to cosmological correlations.
II. Can these corrections become large?'' Phys. Rev. {\bf
D74}(2006) 023508, [hep-th/0605244].

\bibitem{Cogollo:2008bi}
  H.~R.~S.~Cogollo, Y.~Rodriguez and C.~A.~Valenzuela-Toledo,
  ``On the Issue of the $\zeta$ Series Convergence and Loop Corrections in the
  Generation of Observable Primordial Non-Gaussianity in Slow-Roll Inflation.
  Part I: the Bispectrum,''
  arXiv:0806.1546 [astro-ph];














\end{thebibliography}
\end{document}